\documentclass[12pt]{iopart}

\usepackage{graphicx}
\usepackage{lipsum}
\usepackage{wasysym}
\usepackage{amssymb}
\usepackage{multirow}
\usepackage{hyperref}
\usepackage{soul}

\begin{document}

%%%%%%%%%%%%%%%%%%%%%%%%%%%%%%%%%%%%%%%%%%
\title
[Rapid solidification microstructure in Mg alloys]
{Emergence of rapid solidification microstructure in additive manufacturing of a Magnesium alloy}

\author{
Damien~Tourret$^{1,*}$, 
Rouhollah~Tavakoli$^{1,2}$, 
Adrian~D.~Boccardo$^{1}$, 
Ahmed K.~Boukellal$^{1,3}$, 
Muzi~Li$^{1}$
 and 
Jon~Molina-Aldareguia$^{1,4}$
}

\address{$^1$~IMDEA Materials, Getafe, Spain}
\address{$^2$~Sharif University of Technology, Tehran, Iran}
\address{$^3$~Universit\'e de Lorraine, IJL/CNRS, Nancy, France}
\address{$^4$~Department of Mechanical Engineering, Universidad Politécnica de Madrid, 28006 Madrid, Spain}
\ead{damien.tourret@imdea.org}

\begin{abstract}

Bioresorbable Mg-based alloys with low density, low elastic modulus, and excellent biocompatibility are outstanding candidates for temporary orthopedic implants.
Coincidentally, metal additive manufacturing (AM) is disrupting the biomedical sector by providing fast access to patient-customized implants.
Due to the high cooling rates associated with fusion-based AM techniques, they are often described as rapid solidification processes.
However, conclusive observations or rapid solidification in metal AM --- attested by drastic microstructural changes induced by solute trapping, kinetic undercooling, or morphological transitions of the solid-liquid interface --- are scarce.
Here we study the formation of banded microstructures during laser powder-bed fusion (LPBF) of a biomedical-grade Magnesium-rare earth alloy, combining advanced characterization and state-of-the-art thermal and phase-field modeling.
Our experiments unambiguously identify microstructures as the result of an oscillatory banding instability known from other rapid solidification processes.
Our simulations confirm that LPBF-relevant solidification conditions strongly promote the development of banded microstructures in a Mg-Nd alloy.
Simulations also allow us to peer into the sub-micrometer nanosecond-scale details of the solid-liquid interface evolution giving rise to the distinctive banded patterns.
Since rapidly solidified Mg alloys may exhibit significantly different mechanical and corrosion response compared to their cast counterparts, the ability to predict the emergence of rapid solidification microstructures (and to correlate them with local solidification conditions) may open new pathways for the design of bioresorbable orthopedic implants, not only fitted geometrically to each patient, but also optimized with locally-tuned mechanical and corrosion properties.

\end{abstract}

\vspace{2pc}

%%%%%%%%%%%%%%%%%%%%%%%%%%%%%%%%%%%%%%%%%%
\newpage
\section{Introduction}
\label{sec:intro}

Metal additive manufacturing (AM) is already making a transformative impact in the biomedical sector, by enabling on-demand cost- and time-effective fabrication of implants customized to the needs of each patient.
In particular, biodegradable and bioabsorbable metals -- like magnesium (Mg), iron (Fe), and zinc (Zn) -- can degrade gradually {\it in vivo} and be either absorbed or eliminated by the body as the surrounding tissue heals, hence circumventing the need for an extra removal surgery \cite{liu2019fundamental, qin2019additive}.
Among these materials, Mg alloys, often combined with rare earth (RE) elements, constitute a prime candidate for orthopedic implants \cite{witte2010history, lee2016long, zhao2017current}, as they offer good biocompatibility, osteopromotive properties \cite{laires2004role, de2015magnesium}, and mechanical properties close to that of human bones \cite{staiger2006magnesium}, thus reducing stress shielding during loading of the bone-implant interface \cite{zhao2017current}. 
The cytotoxicity of rare earth (RE) elements has been studied both {\it in vitro} and {\it in vivo} \cite{feyerabend2010evaluation, willbold2015effect, liu2019mechanical}. Some of them, such as lanthanum (La), cerium (Ce) and praseodymium (Pr) exhibit high cytotoxicity --- in particular, Ce and Pr chlorides present severe hepatotoxicity \cite{nakamura1997differences}. In terms of 50\% lethal dose (LD$_{50}$) \cite{feyerabend2010evaluation}, Ce shows the highest cytotoxicity (10~mg/kg), followed by La (150~mg/kg), but several other RE elements ($> 550~$mg/kg for Gd, Nd, Dy, and Y) are viable for biomedical applications. For instance, Mg--1.8Zn--0.2Gd alloy was found to exhibit little to no toxicity on several different cell lines \cite{bian2018vitro}, and Nd was found to improve mechanical properties and corrosion resistance within acceptable toxicity levels \cite{zhang2012microstructure}.
Moreover, high cooling rates were reported to be efficient in modifying microstructures in a wide range of Mg alloys, resulting in improved mechanical properties \cite{gjestland1991stress, srivatsan1995tensile, zhang2013improving} and corrosion resistance \cite{liao2012improved,zhang2013improving, willbold2013biocompatibility, shuai2017laser}.
Therefore, fusion-based AM processes, e.g. laser powder-bed fusion (LPBF), which usually yield high cooling rates, provide a promising route to tune processing parameters and achieve desired microstructures, properties, and performance of 3D-printed Mg-based bioimplants.
Yet, the fundamental understanding of microstructure formation in far-from-equilibrium conditions remains incomplete.

In solidification theory \cite{kurz2023fundamentals, dantzig2016solidification}, {\it rapid solidification} refers to a regime in which the solid-liquid interface exhibits a strong departure from equilibrium when reaching a sufficiently high velocity $V$. 
One manifestation of rapid solidification is {\it solute trapping}, whereby the solid phase grows with a solute concentration in excess of its equilibrium solubility limit.
Thus, the solute partition coefficient $k=c_s/c_l$, with $c_s$ and $c_l$ the solute concentration on the solid and liquid sides of the interface (respectively), deviates from its equilibrium value and tends toward unity as $V$ approaches and surpasses a characteristic diffusion velocity $V_{\rm d}$, typically of the order of 1\,m/s.
Meanwhile, both liquidus and solidus lines of the phase diagram converge toward the so-called $T_0-$line as $V$ increases.
The velocity-dependence of the interface solute partition coefficient, $k$, and of the liquidus slope, $m$, are commonly described using the continuous growth model (CGM) \cite{aziz1982model, aziz1988continuous, aziz1994transition}
\begin{eqnarray}
\label{eq:cgm:kv}
k(V) & = \frac{k_{\rm e}+V/V_{\rm d}}{1+V/V_{\rm d}} \\
\label{eq:cgm:mv}
\frac{m(V)}{m_{\rm e}} &= \frac{1-k(V)+[k(V)+(1-k(V))\alpha]\ln[k(V)/k_{\rm e}]}{1-k_{\rm e}} 
\quad,
\end{eqnarray}
where $k_{\rm e}$ and $m_{\rm e}$ stand for the respective equilibrium ($V\ll V_{\rm d}$) values of $k$ and $m$, and $\alpha$ is a solute drag coefficient.
Another rapid solidification effect is the occurrence of {\it kinetic undercooling}, associated with a deviation of the interface temperature, $T$, proportional to the interface velocity. 
The resulting velocity-dependent liquidus and solidus temperatures (for a planar interface, i.e. without curvature undercooling) are thus given by 
\begin{eqnarray}
\label{eq:TlV}
T_{\rm L}(V) &= T_{\rm M} - m(V)c_\infty - \frac{V}{\mu} \\
\label{eq:TsV}
T_{\rm S}(V) &= T_{\rm M} - \frac{m(V)}{k(V)}c_\infty - \frac{V}{\mu}
\end{eqnarray}
with $T_{\rm M}$ the melting temperature of the solvent species, $c_\infty$ the alloy nominal solute concentration, and $\mu$ the interface kinetic coefficient.
While solute trapping can be probed by a compositional analysis of rapidly solidified samples, kinetic undercooling is much harder to measure experimentally, since it requires correlating the interface velocity with its temperature during solidification under deep undercooling conditions \cite{willnecker1989evidence, lum1996high}.
Finally, the most evident observable indication of rapid solidification is the transition of the microstructure pattern from dendrites to a planar interface, at velocities beyond the {\it absolute stability} threshold $V_{\rm a}$.
An extension of the linear perturbation theory by Mullins and Sekerka \cite{mullins1964stability} to high undercooling \cite{trivedi1986morphological} leads to the implicit definition of the threshold velocity above which the planar interface is stable
\begin{equation}
\label{eq:absstab}
V_{\rm a}=\frac{D_{\rm L}\Delta T_0(V_{\rm a})}{k(V_{\rm a})\Gamma}=\frac{D_{\rm L}m(V_{\rm a})c_\infty[1-k(V_{\rm a})]}{k(V_{\rm a})^2\Gamma}
\end{equation}
with $D_{\rm L}$ the solute diffusion coefficient in the liquid, $\Delta T_0$ the alloy freezing range between the liquidus ($T_{\rm L}$) and solidus ($T_{\rm S}$) temperatures, and $\Gamma$ the Gibbs-Thomson coefficient of the solid-liquid interface. 
Equation \eref{eq:absstab} only provides an order of magnitude of the absolute stability limit, which is better estimated by the maximum of $T_{\rm S}(V)$ from \eref{eq:TsV} \cite{carrard1992banded, karma1992dynamics, karma1993interface, kurz1996banded, tourret2023morphological}, while the region with $\rmd T/\rmd V>0$ may lead to an oscillatory instability at the origin of {\it banded microstructures} \cite{kurz1996banded}.

Indeed, intermediate velocities between the dendritic ($V\ll V_{\rm d}$) and planar partitionless ($V\gg V_{\rm d}$) regimes may lead to an oscillatory instability of the solid-liquid interface. 
Resulting microstructures consist of an alternation of (i) featureless regions of homogeneous supersaturated solute content with (ii) patterned regions exhibiting solute segregation and often decorated with small particles of secondary phases such as intermetallic precipitates.
The resulting bands are nearly parallel to the solidification front (i.e. normal to the main temperature gradient and growth direction).
These banded microstructures have been identified in rapidly solidified samples using a broad range of processing techniques (e.g. splat quenching, melt spinning, laser melting, etc.), notably on several of Al-based alloys (e.g. Al-Cu \cite{williams1977microstructural, zimmermann1991characterization,carrard1992banded}, Al-Pd \cite{sastry1981metastable}, Al-Zr \cite{pandey1986microstructural}, Al-Fe \cite{gremaud1991banding},  Al-Er \cite{gianoglio2020banded}, etc.) but also other alloys, like Ag-Cu \cite{boettinger1984effect}.
These observations have shown that the width of the patterned portion of the bands decreases with an increase in growth rate \cite{gremaud1991banding}.
Recently, the formation of these bands in laser-melted thin Al-Cu samples was also observed {\it in situ} within a transmission electron microscope (TEM), allowing the measurements and correlation of the morphological transition with an increase of the interface velocity \cite{mckeown2014situ,mckeown2016time}.

Analytical and numerical studies have identified the origin of the oscillatory instability as a subtle combination of interface undercooling and solute trapping \cite{carrard1992banded, karma1992dynamics, karma1993interface}. 
Essentially, within a velocity range just below absolute planar stability, the effect of solute trapping is dominant compared to the effect of atomic attachment kinetics, resulting in an increase in solidus temperature (i.e. the stable temperature for a planar interface) with interface velocity ($dT/dV>0$). Within a temperature gradient moving at a given velocity $V$, this range of velocity is understood to be unstable for the growth of a planar interface. Hence, therein, the interface oscillates between two stable branches of solutions with $dT/dV<0$, namely corresponding to a dendritic pattern at relatively lower $V$ and a planar interface pattern at relatively higher $V$. This oscillation cycle, first theorized considering instantaneous transition between the two stable branches \cite{carrard1992banded} was later confirmed by further analytical and numerical studies \cite{karma1992dynamics, karma1993interface}.
These studies have highlighted the key role of the latent heat diffusion in the selection of the band spacing --- rendering it essentially unaffected by the temperature gradient \cite{karma1992dynamics,karma1993interface}, while bands are inversely proportional to the temperature gradient if latent heat is neglected \cite{carrard1992banded}.
Recent phase-field simulations have reproduced quantitatively the banding mechanism observed {\it in situ} in Al-Cu TEM samples \cite{ji2023microstructural} and shown that the onset of banding instability was closely approximated by \eref{eq:absstab}, in spite of the underlying linear stability analysis not incorporating anisotropies of excess free energy and kinetic coefficient essential to the interface pattern selection \cite{langer1986solvability, brener1991pattern, bragard2002linking}.

In this context, although LPBF is commonly referred to as a ``rapid solidification'' process, whether or not the solid-liquid interface strongly departs from equilibrium is often ambiguous.
Most additively manufactured microstructures indeed exhibit patterns (dendrites) as well as significant intergranular and interdendritic solute segregation \cite{herzog2016additive,debroy18}, indicative of a relatively weak departure from equilibrium at the solid-liquid interface.
Rather, ``banded'' microstructures reported in metal AM often stem from a periodic change in local processing conditions, for instance due to gas jets disturbing the melt pool and periodically changing the local cooling rate \cite{hwa2021microstructural} or post-solidification microstructure evolution in the heat affected zone \cite{ho2019origin}.

Interestingly, nearly forty years ago, Mg alloys processed by rapid solidification processes, such as splat quenching and melt spinning, were reported to contain ``featureless zones'' indicative of partitionless planar solidification beyond absolute stability \cite{flemings1984rapid}. 
These featureless regions exhibited a greater hardness than both the as-received (primary phase) alloys and the dendritic regions of the rapidly solidified material. 
More recently \cite{li2021microstructure}, a bioresorbable Mg alloy (WE43) additively manufactured via LPBF was reported to exhibit intragranular banded microstructures reminiscent of those observed in earlier studies of rapid solidification microstructures in Al alloys.
The underlying objective of the present article is to investigate, using state-of-the-art quantitative modeling tools, whether solidification conditions relevant to LPBF of Mg alloy WE43 are indeed expected to result in the oscillatory banding instability linked to a strong equilibrium departure at the solid-liquid interface. 
To do so, we apply a recently proposed quantitative phase-field model of rapid solidification \cite{ji2023microstructural}, using local solidification conditions estimated by a thermal model of laser melting and solidification of a powder bed. 
We explore microstructural patterns expected in different regions of the melt pool, as well as their complex formation dynamics.

%%%%%%%%%%%%%%%%%%%%%%%%%%%%%%%%%%%%%%%%%%
\section{Methods}
\label{sec:methods}

%%%%%%%%%%%%%
\subsection{Experiments}

We support our analyses with experimental data already presented in detail elsewhere \cite{li2021microstructure}, including electron microscopy, X-ray tomography, mechanical and corrosion testing, as well as {\it in vitro} cytocompatibility. 
Hence, here we only summarize the key information relevant to the material and manufacturing conditions leading to the emergence of rapid solidification microstructures and the methods used in their characterization.
The reader is referred to earlier works for further relevant details on additive manufacturing processing \cite{kopp2019influence} and sample characterization and testing \cite{li2021microstructure}.

\subsubsection{Material and manufacturing.} %%%%%%%%%%%%%

The material feedstock is a powder of Mg alloy WE43 MEO (Meotec GmbH, Aachen, Germany) with a nominal composition of 1.4--4.2\,wt\%\,Y, 2.5--3.5\,wt\%\,Nd, $<1\,$wt\%\,(Al, Fe, Cu, Ni, Mn, Zn, Zr), and balance Mg  with a particle distribution D$_{10} = 29.1\,$\textmu m, D$_{50} = 45.8\,$\textmu m and D$_{90} = 64.4\,$\textmu m \cite{kopp2019influence}, naturally covered with a thin Y$_2$O$_3$ passivation layer limiting its flammability. 
Lattice structures (open scaffolds) of size (10~mm)$^3$, with a body-centered cubic unit cell structure and different strut diameters (250 to 750~\textmu m), were additively manufactured by LPBF in an Ar atmosphere with a laser power $P\approx90~$W, a beam diameter $\diameter=90~$\textmu m, a layer thickness $H_{\rm p}\approx30~$\textmu m, and a $0^\circ/90^\circ$ scan strategy across successive layers.
In order to prepare TEM samples, an additional bulk cylindrical sample (8~mm in diameter, 30~mm in length) was manufactured with similar processing parameters as the lattices.

\subsubsection{Microstructural characterization.} %%%%%%%%%%%%%

For scanning electron microscope (SEM) observation, samples were cut out of the scaffolds, then manually ground with SiC abrasive paper 4000 grit, followed by mechanical polishing with 3\,\textmu m and 0.25\,\textmu m diamond paste using an alcohol-based lubricant to avoid oxidation, and finally etching with a mixture of 75\,ml ethylene glycol, 24\,ml distilled water and 1\,ml nitric acid.
For TEM analysis, thin samples of 200\,\textmu m thickness were cut from the cylindrical specimen, manually polished down to about $80$\,\textmu m, and then polished with an etchant of 10.6\,g lithium chloride, 22.32\,g magnesium perchlorate, 200\,ml 2-butoxy-ethanol and 1000\,ml methanol at $-40^\circ$C and 95\,V. 
TEM observation was carried out in a FEI Talos equipped with a field emission gun operating at 200\,kV. 
Chemical composition mapping via energy-dispersive X-ray spectroscopy (EDS) was performed under scanning-transmission electron microscopy (STEM) mode with a SuperX detector, with an acquisition time between 30 and 60\,min, and the resulting data was analyzed using Bruker QUANTAX software.
The micrographs appearing in section~\ref{sec:results} (figure~\ref{Fig:SEM}) were further processed (brightness and contrast adjustment) using the software ImageJ \cite{schneider2012nih}.

%%%%%%%%%%%%%
\subsection{Thermal modeling}
\label{sec:meth:thermal}

\subsubsection{Finite element model.} %%%%%%%%%%%%%
We assessed the temperature profile using the finite element (FE) software Abaqus \cite{smith_2009}.
The main objective of these calculations is to estimate the typical melt pool dimensions, and the temperature gradient $G$ and growth velocity $V$ experienced by the solidification front.
We do not explore the effect of geometry of the printed lattice structures, but rather seek an order of magnitude of the  $(G,V)$ conditions to use in the phase-field simulations of microstructure formation, and hence simply simulate a single linear melting/solidification track.

The heat transfer equation, prescribing the evolution of the temperature field $T$ with time $t$, follows \cite{urrutia2014thermal,Ning_2021}
\begin{equation}
  \rho c_{p} \frac{\partial T}{\partial t}= \nabla\cdot \left( \kappa \,\nabla T \right)+Q_{\rm L}+Q_\phi \quad,
\end{equation}
where $\rho$ is the density, $c_{p}$ is the specific heat capacity, $\kappa$ is the (isotropic) conductivity, $Q_{\rm L}$ is the absorbed heat due to
the laser, and $Q_\phi$ is the absorbed/released heat due to phase transformations.
The volumetric heat source from the moving laser is introduced via the \texttt{dflux} subroutine \cite{cheng_2014} as 
\begin{equation}
  Q_{\rm L}=\alpha_{\rm p} \frac{G_{\rm L}I_{z}}{H_{\rm L}}
\end{equation}
where $\alpha_{\rm p}$ is the absorptivity of the powder bed and $H_{\rm L}$ is the laser penetration depth. 
The Gaussian distribution along the top $(z=z_{\rm L})$ surface, $G_{\rm L}$, and the parabolic decay along the laser penetration direction $z$, $I_z$, are given by  
\begin{eqnarray}
  G_{\rm L} &= \frac{2 P}{\pi \diameter^{2}} \exp \left[ -2\frac{(x-x_{\rm L})^{2}+(y-y_{\rm L})^{2}}{\diameter^{2}} \right] \\
  I_z &= \frac{1}{0.75} \left[ -2.25 \left(\frac{z_{\rm L}-z}{H_{\rm L}} \right)^{2}+1.5  \left(\frac{z_{\rm L}-z}{H_{\rm L}} \right) +0.75 \right] \quad,
\end{eqnarray}
where $P$ is the laser power, $\diameter$ is the laser diameter, and ($x_{\rm L}, y_{\rm L}, z_{\rm L})$ is the location of the laser heat source center 
within a cartesian coordinate system $(x, y, z)$, with the laser moving at a velocity $V_{\rm L}$ in the $x+$ direction.
The source term related to phase transformation during melting and solidification of the alloy is implemented via the \texttt{hetval} subroutine and calculated as
\begin{equation}
  Q_\phi=\rho L_{\rm f} \frac{\partial f_{\rm s}}{\partial t}\quad,
\end{equation}
where $L_{\rm f}$ is the latent heat of fusion and $f_{\rm s}$ the volume fraction of solid, approximated by a linear interpolation between the alloy liquidus ($T_{\rm L}$) and solidus ($T_{\rm S}$) temperatures.
We apply boundary conditions (BCs) following Newton's law
\begin{equation}
  q_{\rm c}=-h_{\rm c}(T-T_{\infty}) \quad,
\end{equation}
where $q_{\rm c}$ is the normal heat flux, $h_{\rm c}$ is the interfacial heat transfer coefficient, and $T_{\infty}$ is the surrounding temperature. 
This condition is applied on all domain boundaries, but the domain is taken large enough laterally to result in a negligible effect of BCs in $x$ and $y$ directions on the melt pool dimensions, while different heat transfer coefficients are applied on the top ($z+$) alloy/air interface and the bottom ($z-$) alloy/build plate interface.
Different properties are considered for the powder bed and for the dense (solid or liquid) material. 
The initial powder-bed state transitions to the liquid state when the local temperature exceeds the alloy liquidus temperature, $T_{\rm L}$, and is then considered solid when the temperature drops below the solidus temperature, $T_{\rm S}$.

\subsubsection{Parameters.} %%%%%%%%%%%%%
Material properties and process parameters are listed in Tables~\ref{tab:fe:material} and \ref{tab:fe:process}, respectively.
Most thermal properties (Table~\ref{tab:fe:material}) are approximated by those of pure Mg using values from classical handbooks \cite{valencia2008asm,brandes2013smithells}, including temperature-dependent conductivity $\kappa(T)$ and specific heat capacity $c_p(T)$ \cite{brandes2013smithells} --- the latter being converted from molar to mass using the molar mass $M_{\rm Mg}=24.305$~g/mol.
For consistency with simulations of microstructure formation, liquidus and solidus temperature are calculated for a Mg-3wt\%Nd alloy (see section~\ref{sec:meth:pf}).
The powder bed is assumed to have a slightly reduced density (by a factor 0.7, just below that of a close packing of identical spheres, i.e. with a porosity fraction $\xi=0.3$) but a significantly reduced conductivity.
For the latter, we follow the model proposed by Sih and Barlow \cite{sih2004prediction}, derived from the Zehner-Schl\"under-Damk\"ohler equation \cite{damkohler1936einflusse, zehner1970warmeleitfahigkeit} assuming spherical particles 
\begin{eqnarray}
\fl       \frac{\kappa_{\rm p}}{\kappa_{\rm g}} = &\left(1-\sqrt{1-\xi}\right) \left(1+\xi \frac{\kappa_{\rm r}}{\kappa_{\rm g}}\right) %\nonumber\\
        + \sqrt{1-\xi} \left\{\frac{2}{1-\frac{\kappa_{\rm g}}{\kappa}} \left[\frac{1}{1-\frac{\kappa_{\rm g}}{\kappa}} \ln\left(\frac{\kappa}{\kappa_{\rm g}}\right) - 1\right] + \frac{\kappa_{\rm r}}{\kappa_{\rm g}}\right\}
    \label{eq:powderkappa}
\end{eqnarray}
where $\kappa$, $\kappa_{\rm p}$, and $\kappa_{\rm g}$ are the conductivity of the bulk (dense) material, powder bed, and gas, respectively.
The thermal conductivity due to radiation among particles, $\kappa_{\rm r}$, is
\begin{equation}
       \kappa_{\rm r} = 4Fd\sigma T^3 
    \label{eq8}
\end{equation}
where $F$ is a view factor, $d$ is the average diameter of powder particles, and $\sigma$ is the Stefan-Boltzmann constant ($5.67\times10^{-8}$~W/m$^2$/K$^4$).
Considering the temperature-dependent thermal conductivity of Mg from \cite{brandes2013smithells} (Table~\ref{tab:fe:material}) and that of Ar gas measured in \cite{chen1975thermal}, a view factor $F=1/3$ \cite{damkohler1936einflusse, sih2004prediction}, and a particle size $d\approx 50~$\textmu m, we found that the conductivity of the powder-bed was well approximated by a second order polynomial of the temperature and used directly this approximation, listed in Table~\ref{tab:fe:material}, in the simulations.
The resulting conductivity of the powder bed is over two orders of magnitude lower than that of the bulk material, with $\kappa_{\rm p}/\kappa\approx0.0058$ at 923~K.
Process parameters (Table~\ref{tab:fe:process}) such as laser parameters ($P$, $V_{\rm L}$, $\diameter$) and the powder layer thickness ($H_{\rm p}$) are taken according to experimental conditions.
The laser penetration depth $H_{\rm L}\approx1.5H_{\rm p}$ and top/bottom heat transfer coefficients are reasonable orders of magnitude typically used in metal processing simulations~\cite{urrutia2014thermal,Ning_2021}.

%%%%%%%%%%%%%%%%%%%%%
\begin{table}[t]
\caption{Material properties considered in the thermal simulation ($T$ in K).}
\label{tab:fe:material}
\centering
\footnotesize
\begin{tabular}{l c c r l c}
\br
Parameter & Region & Symbol & Value & Unit & Source \\
\mr
Density & Bulk & $\rho$ & 1590 & kg/m$^3$ & \cite{valencia2008asm} \\
Density & Powder bed & $\rho_{\rm p}$ & $(1-\xi)\rho$ & &  \\
Thermal conductivity & Bulk & $\kappa$ & \multicolumn{2}{c}{see below $^{a}$} & \cite{brandes2013smithells}\\
Thermal conductivity & Powder bed & $\kappa_{\rm p}$ & \multicolumn{2}{c}{see below $^{b}$} &  \\
Specific heat capacity & Bulk \& Powder & $c_p$ & \multicolumn{2}{c}{see below $^{c}$} & \cite{brandes2013smithells}\\
Latent heat of fusion & Bulk \& Powder & $L_{\rm f}$ & 349 & J/g & \cite{valencia2008asm}\\
Liquidus temperature & Bulk \& Powder & $T_{\rm L}$ & 912.7 & K & \cite{okamoto2016alloy}\\
Solidus temperature & Bulk \& Powder & $T_{\rm S}$ & 859.7 & K & \cite{okamoto2016alloy}\\
Absorptivity & Powder bed & $\alpha_{\rm p}$ & 0.6 & & \\
Porosity fraction & Powder bed & $\xi$ & 0.3 & &  \\
\br
\end{tabular}\\
\begin{tabular}{l l r r r r r r r r l}
$^{a}$ & $T$ & 293 & 373 & 473 & 673 & 923 & 973 & 1073 & 1273 & K \\
& $\kappa(T)$ &167 & 167 & 163 & 130 & 78 & 81 & 88 & 100 & W/m/K \\
\br
\end{tabular}\\
\begin{tabular}{l l}
$^{b}$ & $\kappa_{\rm p} \approx (0.1118+5.90\times10^{-4}\,T-2.47\times10^{-7}\,T^2)$ W/m/K\\
\br
\end{tabular}\\
\begin{tabular}{l l}
$^{c}$ & $c_p(T)=172.26\,(a+b\times10^{-3}\,T+c\times10^5\,T^{-2})$ J/kg/K \\
&\quad with $a=5.33$, $b=2.45$, and $c=-0.103$ in the solid\\
&\quad and $a=7.80$ and $b=c=0$ in the liquid\\
\br
\end{tabular}
\end{table}
\normalsize
%%%%%%%%%%%%%%%%%%%%%

%%%%%%%%%%%%%%%%%%%%%
\begin{table}[t]
\caption{Process parameters considered in the thermal simulation.}
\label{tab:fe:process}
\centering
\footnotesize
\begin{tabular}{l c r l}
\br
Parameter & Symbol & Value & Unit \\
\mr
Laser power & $P$ & 90 & W \\
Laser diameter & $\diameter$ & 90 & \textmu m \\
Laser velocity & $V_{\rm L}$ & 1.0 & m/s \\
Laser penetration depth & $H_{\rm L}$ & 45 & \textmu m \\
Powder layer thickness & $H_{\rm p}$ & 30 & \textmu m \\
Surrounding temperature & $T_\infty$ & 298 & K \\
Heat transfer coefficient (top) & $h_{\rm c}^{z+}$ & 50 & W/m$^{2}$/K  \\
Heat transfer coefficient (bottom) & $h_{\rm c}^{z-}$ & 500 & W/m$^{2}$/K \\
\br
\end{tabular}
\end{table}
\normalsize
%%%%%%%%%%%%%%%%%%%%%

\subsubsection{Simulations.} %%%%%%%%%%%%%

Since we only seek reasonable estimates of the $(G,V)$ conditions in the region surrounding the melt pool during LPBF of our Mg alloy, here we chose to disregard the particular geometry of the lattices and the printing strategy, and rather calculate the temperature distribution along a straight single track at steady state using relevant material properties (Table~\ref{tab:fe:material}) and processing parameters (Table~\ref{tab:fe:process}).
The simulated sample consists of a wide plate of solid Mg alloy (4.5~mm long, 3.75~mm wide, and 470~\textmu m high) with a $H_{\rm p}=30$~\textmu m layer of Mg alloy powder at the top. 
The domain is divided in two regions: (i) the region directly affected by the laser (105~\textmu m deep and 750~\textmu m wide, laterally centered on the laser track path) finely meshed with 1\,902\,600 linear hexahedral elements (DC3D8) of edge size 2~\textmu m, and (ii) the overall surrounding region meshed with 224\,977 linear tetrahedral elements (DC3D4). 
We defined the mesh size by performing a convergence analysis of the temperature field in the melt pool.
The simulated time is 0.9~ms, which is sufficient to reach a stable steady melt pool shape and temperature profile therein.
The simulation completion took approximately 9~hours on 28 CPU cores (Intel Xeon Gold 6130 2.10GHz).

\subsubsection{Post-processing.} %%%%%%%%%%%%%

From the FE-calculated thermal field, we extract local $(G,V)$ solidification conditions along the tail of the melt pool, to be used in the one-dimensional approximation of the temperature field in the PF simulations. 
Since the melt pool is nearly elliptical in shape (see section~\ref{sec:results}, figure~\ref{Fig:FE}), we approximate the location of $T_{\rm L}$ and $T_{\rm S}$ isotherms by ellipses \cite{elahi2022multiscale} with  
\begin{equation}
r_{\rm L}(\theta) = \sqrt{\frac{(l_{\rm L} d_{\rm L})^2} { \big(d_{\rm L} \cos
(\theta)\big)^2 + \big( l_{\rm L}\sin (\theta)\big)^2}}
\label{eq:rl}\\ %
\end{equation}
and
\begin{equation}
r_{\rm S}(\theta)  = \sqrt{\frac{(l_{\rm S} d_{\rm S})^2} { \big(d_{\rm S} \cos
(\theta)\big)^2 + \big( l_{\rm S}\sin (\theta)\big)^2}}
\quad,
\label{eq:rs}
\end{equation}
where $r_{\rm L}$ and $r_{\rm S}$ are the respective radii of the $(T=T_{\rm L})$ and $(T=T_{\rm S})$ ellipses as a function of the angle measured counterclockwise from the top surface \cite{elahi2022multiscale}
\begin{equation}
\theta = \tan^{-1} \bigg|\frac{z-z_0}{x-x_0}\bigg|
\label{eq:theta}
\end{equation}
with $(x_0,z_0)$ the assumed common center of both ellipses (close but not exactly overlapped with $(x_{\rm L},z_{\rm L})$).
The dimensions of the melt pool tail appear explicitly in \eref{eq:rl} and \eref{eq:rs} as the length ($l_{\rm L}$, $l_{\rm S}$) and depth ($d_{\rm L}$, $d_{\rm S}$) of the solidus (subscript S) and liquidus (L) isotherms.
A linear radial interpolation of the temperature between $T_{\rm L}$ and $T_{\rm S}$ provides a reasonable approximation of the temperature field in the region where solidification takes place \cite{elahi2022multiscale}.
Hence, since the polar component of the temperature gradient, $(1/r)\partial T/\partial\theta$, is usually much lower than its radial component, this approximation also provides a simple and convenient approximation of the amplitude of the temperature gradient 
\begin{equation}
G(\theta) \approx \frac{\partial T}{\partial r} \approx \frac{T_{\rm L}-T_{\rm S}}{r_{\rm S}(\theta)-r_{\rm L}(\theta)} \quad,
\label{eq:Gellip}
\end{equation}
whereas the local velocity of the isotherms, i.e. the local solidification velocity, is simply 
\begin{equation}
V(\theta) = V_{\rm L} \cos(\theta)
\quad.
\label{eq:Vellip}
\end{equation}

%%%%%%%%%%%%%
\subsection{Microstructure formation modeling}
\label{sec:meth:pf}

\subsubsection{Phase-field model.} %%%%%%%%%%%%%
To simulate microstructure formation, we use a phase-field (PF) model for rapid solidification recently proposed by Ji et al. \cite{ji2023microstructural}. 
We adapt the formulation of the anisotropic terms for a hexagonal (hcp) Mg alloy, and apply it in two dimensions (2D), within the basal plane, considering local values of temperature gradient $G$ and growth velocity $V$ estimated via the thermal calculations (Section~\ref{sec:meth:thermal}).

It is known that using a diffuse interface width $W$ much greater than the actual physical width of the solid-liquid interface $W_0$ leads to artificial solute trapping effect, even at moderate growth velocities, which is typically corrected by adding an ``anti-trapping'' term into the solute conservation equation \cite{karma2001phase, echebarria2004quantitative, kim2007phase, ohno2009quantitative}.
Meanwhile, when using a physically realistic interface width $W=W_0\sim 1~$nm, PF models may be capable of quantitatively capturing solute trapping \cite{ahmad1998solute,danilov2006phase,galenko2011solute, steinbach2012phase}.
In the model used here \cite{ji2023microstructural}, the spurious trapping effect due to the use of $W/W_0>1$ is compensated via an increased solute diffusivity through the interface.
For a given value of $S=W/W_0$, the interpolation function for the diffusivity across the interface can be chosen to ensure that results for $S>1$ remain quantitatively close to those at $S=1$ in terms of partition coefficient $k(V)$ and liquidus slope $m(V)$ across a broad range of interface velocity $V$, where $k$ varies from its equilibrium value $k_{\rm e}$ at moderate $V$ up to $k\to 1$ at high $V$.
By enabling upscaled yet quantitative simulations with $S>1$, the resulting PF simulations allow studying the morphological details of solid-liquid interface pattern evolution at experimentally relevant length and time scales. For instance, the model was recently used to quantitatively reproduce the banding growth instability observed in thin Al-Cu samples melted and observed inside a transmission electron microscope \cite{ji2023microstructural,mckeown2014situ,mckeown2016time}.

The evolution of the phase field, $\phi$, and solute concentration field, $c$, follows \cite{ji2023microstructural}
\begin{eqnarray}
\label{eq:dpdt}
\tau(\mathbf{n})\frac{\partial \phi}{\partial t} &= &
\vec\nabla \cdot\left[ W(\mathbf{n})^2\vec\nabla \phi\right] +\phi-\phi^3
+\sum_{\eta=x,y}\left[ \partial_\eta \left( |\vec\nabla\phi|^2 W(\mathbf{n}) \frac{\partial W(\mathbf{n})}{\partial (\partial_\eta\phi)}\right) \right] 
\nonumber \\
&& -\lambda g'(\phi)\left[ c+\frac{T-T_{\rm M}}{m_{\rm e}} \exp\left\{b(1+g(\phi))\right\} \right] ~,\\
\label{eq:dcdt}
\quad\quad \frac{\partial c}{\partial t} &= & \vec\nabla\cdot \left\{ D_{\rm L} q(\phi)c\vec\nabla[\ln c - bg(\phi)] \right\} ~,
\end{eqnarray}
where we use $\partial_\eta$ to denote $\partial/\partial\eta$, $b=\ln(k_{\rm e})/2$, $g(\phi)=15(\phi-2\phi^3/3+\phi^5/5)/8$, and $\lambda=(2\sqrt{2}/3)bm_{\rm e}SW_0/[\Gamma(k_{\rm e}-1)]$, with the Gibbs-Thomson coefficient $\Gamma=\gamma_0T_{\rm M}/(\rho L_{\rm f})$, where $\gamma_0$ is the solid-liquid interface (average) excess free energy, $T_{\rm M}$ is the solvent (Mg) melting point, and $\rho L_{\rm f}$ is the latent heat of fusion per unit volume. 
The $(x,y)$ cartesian coordinate system used here differs from the $(x,y,z)$ coordinates used in the FE thermal simulations, as the $x+$ direction of the PF simulation corresponds to the main temperature gradient direction along the fusion line, directed toward the center of the melt pool.
Given a solute diffusivity $D=0$ in the solid and $D=D_{\rm L}$ in the liquid phase, we consider an enhanced diffusivity through the interface, $D(\phi)=D_{\rm L}q(\phi)$, using the same function $q(\phi)=A(1-\phi)/2-(A-1)(1-\phi)^2/4$ as in \cite{ji2023microstructural}, in order to compensate for the spurious solute trapping effect due to $S=W/W_0>1$.
For a given value of $S$, an appropriate value of the coefficient $A$ is estimated similarly as in \cite{ji2023microstructural}, so as to minimize the departure of $k(V)$ and $m(V)$ from curves calculated at $S=1$ across the entire velocity range (see section~\ref{sec:meth:pf:simu}).
We consider the usual frozen temperature approximation, whereby a one-dimensional temperature field is imposed as $T=T_0+G(x-x_0-Vt)$ with $V$ the velocity of the isotherms (i.e. the growth velocity in the $x$ direction at steady state) with respect to a fixed reference temperature $T_0$ at a point $x_0$ at a time $t=0$.
For the hexagonal close-packed Mg alloy, the anisotropic solid-liquid interface excess free energy, $\gamma({\bf n})$, and kinetic coefficient, $\mu({\bf n})$, here considered within the basal plane, are expressed using spherical harmonics \cite{hoyt2003atomistic, sun2006crystal} as
\begin{eqnarray}
\label{eq:aniso_g}
a_{\rm s}({\bf n})&=\frac{\gamma(\bf{n})}{\gamma_0} =  1 - \frac{\varepsilon^\gamma_{20}}{4}\sqrt{\frac{5}{\pi}} + \frac{\varepsilon^\gamma_{66}}{64}\sqrt{\frac{6006}{\pi}} \cos(6\theta) \quad, \\
\label{eq:aniso_k}
a_{\rm k}({\bf n})&=\frac{\mu(\bf{n})}{\mu_0} =  1 - \frac{\varepsilon^\mu_{20}}{4}\sqrt{\frac{5}{\pi}} + \frac{\varepsilon^\mu_{66}}{64}\sqrt{\frac{6006}{\pi}} \cos(6\theta) \quad,
\end{eqnarray} 
where $\gamma_0$ and $\mu_0$ are the average values of $\gamma$ and $\mu$.
In \eref{eq:aniso_g}-\eref{eq:aniso_k}, we use the same form for $\gamma(\bf{n})$ and $\mu(\bf{n})$ and consider that coefficients $\varepsilon^\gamma_{40}=\varepsilon^\gamma_{60}=\varepsilon^\mu_{40}=\varepsilon^\mu_{60}=0$, which was established by molecular dynamics (MD) calculations for $\gamma({\bf n})$ \cite{sun2006crystal} and is here assumed to be also valid for $\mu({\bf n})$.
The resulting anisotropic interface width and relaxation time are $W(\mathbf{n}) = SW_0a_{\rm s}({\bf n})$ and $\tau(\mathbf{n}) = \tau_0 a_{\rm s}({\bf n})^2/a_{\rm k}({\bf n})$, where $\tau_0 = (SW_0)^2/(\Gamma \mu_0)$ and the orientation $\theta$ of the interface normal ${\bf n}=(n_x,n_y)$ is $\theta=n_y/n_x=\tan^{-1}(\partial_y \phi / \partial_x \phi )$.

\subsubsection{Computational implementation.} %%%%%%%%%%%%%
Equations \eref{eq:dpdt}-\eref{eq:dcdt} are solved using an explicit Euler time-stepping scheme with a finite difference spatial discretization, implemented in C-based CUDA (Compute Unified Device Architecture) programming language for massive parallelization on multiple Graphics Processing Units (GPUs).
To minimize artificial grid anisotropy effects, Laplacian and divergence terms in \eref{eq:dpdt}-\eref{eq:dcdt} are discretized using operators that are isotropic at order $h^2$ of the grid spacing $h$ \cite{ji2022isotropic}.
Anisotropic terms are solved similarly as in \cite{tourret2015growth} (see Appendices therein), i.e. by expanding the terms in \eref{eq:dpdt} containing $a_{\rm s}({\bf n})$ or its spatial derivatives, and expressing them as a function of first and second order spatial derivatives of the phase field $\phi$.
The expressions used for the sixfold anisotropy considered here are fully developed in the Appendix at the end of this article.
The resulting anisotropy implementation was verified to yield similar results as a different thoroughly validated implementation \cite{boukellal2024} in the low-to-moderate velocity regime (using the model from \cite{echebarria2004quantitative,tourret2015growth}).

The multi-GPU implementation is limited to GPUs hosted on a single node, using a simple layer-wise domain decomposition.
Therein, similarly as in \cite{elahi2022multiscale}, for a parallelization on $n$ GPUs, the domain of total dimension $L_x\times L_y$ is divided in $n$ layers of dimension $L_x\times(L_y/n+2)$, considering an extra halo layer of points at the top and bottom of each sub-domain for data exchange. 
The time loop has a single global kernel call for the calculation of both $\phi$ and $c$ at the next time step.
Indeed, compared to a classical antitrapping formulation \cite{karma2001phase, echebarria2004quantitative, tourret2015growth}, the absence of a term containing $\partial \phi/\partial t$ in the equation for the evolution of $c$ (namely the antitrapping term) allows to solve both \eref{eq:dpdt} and \eref{eq:dcdt} in the same kernel regardless of parallel thread synchronization, hence slightly improving computational efficiency.
After the execution of the main kernel, the halo grid data is updated via direct GPU-GPU memory copy, thus avoiding costly GPU-to-CPU or CPU-to-GPU data transfer.

\subsubsection{Parameters.} %%%%%%%%%%%%%
Material properties and numerical parameters considered in the phase-field simulations are summarized in Table~\ref{tab:pf:params}.

%%%%%%%%%%%%%%%%%%%%%
\begin{table}[b]
\caption{Material and simulation parameters considered in phase-field simulations.}
\label{tab:pf:params}
\centering
\footnotesize
\begin{tabular}{l c r l c}
\br
Parameter & Symbol & Value & Unit & Source \\
\mr
Nominal alloy concentration&$c_\infty$&$3.0$&wt\%\,Nd \\
Solvent (Mg) melting temperature & $T_{\rm M}$ & $923$ & K & \cite{okamoto2016alloy} \\
Equilibrium solute partition coefficient& $k_{\rm e}$ & $0.163$ & - & \cite{okamoto2016alloy} \\
Equilibrium Liquidus slope& $m_{\rm e}$ & $3.44$ & K/wt\%\,Nd & \cite{okamoto2016alloy} \\
Gibbs-Thomson coefficient & $\Gamma$ & $1.5 \times 10^{-7} $ & K.m & \cite{sun2006crystal}\cite{valencia2008asm} \\
Solute diffusivity in the liquid & $D_{\rm L}$&$5.26\times10^{-9}$&m$^2$/s & \cite{roy1988prediction, protopapas1973theory, yang2009molecular, housecroft2008inorganic}\\
Interface kinetic coefficient & $\mu_0$ & $0.644$ & m/s/K & \cite{xia2007molecular}\\
Interface kinetic anisotropy coefficients & $\epsilon^\mu_{66}$ & $0.0845$ & - & \cite{xia2007molecular}\\
 & $\epsilon^\mu_{20}$ & $-0.523$ & - & \cite{xia2007molecular}\\
Interface energy anisotropy coefficients & $\epsilon^\gamma_{66}$ & 0.003 & - & \cite{sun2006crystal} \\
  & $\epsilon^\gamma_{20}$ & $-0.026$ & - & \cite{sun2006crystal} \\
Growth (i.e. isotherms) velocity & $V$ & $[0.1-1.0]$ & m/s &  \\
Temperature gradient & $G$ & $[4.2-10.3]$ & K/\textmu m &  \\ 
Atomic interface thickness & $W_0$ & $1$ & nm &   \\
Diffuse interface thickness & $W$ & 5 & $W_0$ &   \\
Enhanced interface diffusion coefficient & $A$ & 11 & &  \\
Grid spacing & $h$ & $0.8$ & $W$ &  \\
 &  & = 4.0 & nm &  \\
Explicit time step & $\Delta t$ & $0.6\,\tau_0(h/W)^2/4$ & &  \\
 &  & $\approx 2.5\times10^{-11}$ & s &  \\
\br
\end{tabular}
\end{table}
\normalsize
%%%%%%%%%%%%%%%%%%%%%

We approximate the WE43 alloy as a binary Mg-3wt\%Nd alloy. 
This is motivated by the fact that only yttrium and neodymium have nominal concentrations above one percent in mass (namely with $c_{\rm Y}\in[1.4-4.2]$\,wt\% and $c_{\rm Nd}\in[2.5-3.5]$\,wt\% \cite{li2021microstructure}) and the experimental observation of higher Nd segregation levels \cite{li2021microstructure} (see section~\ref{sec:results}, figure~\ref{Fig:EDX}).
In fact, microstructural characterization (see section~\ref{sec:results}) shows that most of the Y is trapped within the protective Y$_2$O$_3$ passivation layer of the powder, which is not melted but only broken down into flakes during LPBF, hence not participating directly in solute partitioning during solidification.
Moreover, while both Y and Nd exhibit a significant extension of their solubility range in Mg under rapid solidification conditions \cite{froes1989rapid,hehmann1990extension}, the maximum equilibrium solubility (at eutectic temperature) of Nd is substantially lower (4.6\,wt\%) than than of Y ($\approx14$\,wt\%) \cite{okamoto2016alloy}, with a lower partition coefficient of the former.
These observations point toward Nd as the most likely species to predominantly participate in the banding instability.
Furthermore, the low nominal alloy concentration also provides confidence in the appropriateness of the PF model, originally intended for relatively dilute binary alloys.

Phase diagram features ($T_{\rm M}$, $k_{\rm e}$, $m_{\rm e}$) are directly taken from the Mg-Nd binary diagram \cite{okamoto2016alloy}.
The Gibbs-Thomson coefficient is calculated as $\Gamma=\gamma_0T_{\rm M}/(\rho L_{\rm f})$, using an interfacial excess free energy $\gamma_0=0.0899~$J/m$^2$ calculated (for pure Mg) via molecular dynamics (MD) calculations \cite{sun2006crystal} and $\rho$ and $L_{\rm f}$ listed in Table~\ref{tab:fe:material} \cite{valencia2008asm}.
The solute (Nd) diffusion coefficient in the liquid state is approximated as $D_{\rm L}\equiv D_{\rm Mg}^{\rm Nd}\approx (r_{\rm Mg}/r_{\rm Nd}) D_{\rm Mg}^{\rm Mg}$ \cite{roy1988prediction}, considering a self-diffusion coefficient $D_{\rm Mg}^{\rm Mg}\approx5.95\times10^{-9}~$m$^2$/s at $T=T_{\rm M}$ (averaged between values calculated in \cite{protopapas1973theory} and \cite{yang2009molecular}) and using Goldschmidt (metallic) radii $r_{\rm Mg}=160~$pm and $r_{\rm Nd}=181~$pm \cite{housecroft2008inorganic}.
Average values and anisotropies of the solid-liquid interface excess free energy, $\gamma(\bf{n})$, and kinetic coefficient, $\mu(\bf{n})$, were calculated by MD for pure Mg \cite{sun2006crystal,xia2007molecular}, here assuming that coefficients $\varepsilon^\mu_{40}$ and $\varepsilon^\mu_{60}$ in the expression of $\mu(\bf{n})$ can be neglected, as was shown to be the case for $\gamma(\bf{n})$ \cite{sun2006crystal}.

Regarding numerical parameters, we consider a diffuse interface width with $S=5$ and a corresponding coefficient $A=11$ assessed by an analysis of $k(V)$ and $m(V)$ curves across the entire velocity range (see section~\ref{sec:meth:pf:simu}). 
The grid spacing is set as $h=0.8W$ to allow a reasonable spatial description of the $\phi$ profile across the interface, and the explicit time step is chosen for stability of the explicit scheme (with a safety factor $0.6$).

\subsubsection{Simulations. \label{sec:meth:pf:simu}}  %%%%%%%%%%%%%
With the PF model and parameters described above, we performed rapid solidification simulations considering $(G,V)$ conditions sampled along the tail (i.e. solidifying) part of the melt pool in the calculated thermal field (section~\ref{sec:meth:thermal}). Before describing the main 2D PF simulations, we comment on the convergence analysis carried out to assess an appropriate value of the interfacial diffusion coefficient $A$ for a given value of $S=W/W_0>1$, following the same procedure as presented in \cite{ji2023microstructural} (see section~A of the Supplemental Material therein).

In order to identify an appropriate value of $A$ for a given $S$, we consider a one-dimensional problem, and compute the $k(V)$ and $m(V)$ curves for the reference case at $S=A=1$ over a broad range of $V$.
As described in \cite{ji2023microstructural}, assuming that the phase field $\phi$ remains close to its theoretical stationary profile $\phi_0(x)=-\tanh[x/(\sqrt{2}W)]$, the corresponding concentration profile $c(x)$ can be obtained by numerical integration of
%%%%%
\begin{equation}
\label{eq:dcdt_1D_steady}
\frac{d c}{d x}=(c_\infty-c)\frac{V}{D_{\rm L}q(\phi)}+b\,c\,\frac{d g(\phi)}{d x} 
\end{equation}
%%%%%
with $c(\pm\infty)=c_\infty$.
For a given $V$, the partition coefficient $k(V)$ can thus be obtained from the ratio of concentrations on the solid ($c_{s}$) and liquid ($c_{l}$) sides of the interface, respectively approximated as the nominal concentration $c_\infty$ and the maximum value of $c(x)$.
The corresponding slope of the liquidus line is then given by the integration of
%%%%%
\begin{equation}
\label{eq:mV_1D_steady}
\frac{m(V)}{m_{\rm e}} = \frac{b}{(1-k_{\rm e})c_{l}} \int_{-\infty}^{+\infty}{g'(\phi)c\frac{d \phi}{d x} \,\rmd x} \quad.
\end{equation}
%%%%%
Once $k(V)$ and $m(V)$ calculated for a given set of $(A,S)$ values, their deviation from the reference case at $A=S=1$, denoted $\delta_k$ and $\delta_m$, are quantified as the (discrete) average of 
$\sqrt{(k_S-k_1)^2}/k_1$ 
and 
$\sqrt{(m_S-m_1)^2}/m_1$, 
respectively, over the calculated velocity range, with $k_1$ and $m_1$ the reference values of $k$ and $m$ calculated with $A=S=1$ and $k_S$ and $m_S$ the values calculated for $S\neq 1$ \cite{ji2023microstructural}.
Here, equations \eref{eq:dcdt_1D_steady} and \eref{eq:mV_1D_steady} are integrated numerically up to machine precision using Matlab library \texttt{Chebfun} \cite{chebfun} for 20 different velocities $V = 0.024\times 1.5^{\{0, 1, \ldots, 19\}}$\,m/s. 
An optimal value of $A$ is then chosen such that the averages of $\delta_k(A)$ and $\delta_m(A)$ are minimized (see \cite{ji2023microstructural} for further details). 

Then, we perform 2D PF simulations of solidification considering a one-dimensional temperature gradient under various $(G,V)$ conditions listed in Table~\ref{tab:param:meltpool}.
These conditions are sampled between the solidus and liquidus temperatures along the tail of the melt pool in the FE-predicted thermal field (see section~\ref{sec:meth:thermal}).
PF simulations start with a planar solid-liquid interface located at the liquidus temperature, with a small random perturbation of amplitude $h$ (i.e. 4\,nm) in the $x$ direction.
The phase field is initialized with the 1D stationary solution $\phi(x)=-\tanh[(x-x_{\rm i})/(\sqrt{2}W)]$ and the solute concentration field with the corresponding equilibrium $c(x)=c_\infty\exp\left\{ b [g(\phi(x))+1]\right\}$.
The most advanced (highest $T$) point along the solid-liquid interface is kept at the same location $x_{\rm i}$ within the domain during the entire simulation by considering a moving frame in the $x$ direction.
The location $x_{\rm i}$ is chosen to leave a sufficient distance to the upper $x$ domain boundary, i.e. a length $L_{\rm liquid}$ that allows for a decaying concentration profile toward $c_\infty$ in the occurrence of solute partitioning at the interface.
The liquid length $L_{\rm liquid}$ ahead of the interface was set to 1\,\textmu m, which represents at least 19 times the characteristic diffusion length $D_{\rm L}/V$ for all cases.
The Mg preferred growth direction $\langle11\bar20\rangle$ is aligned with the temperature gradient direction ($x+$).
The domain size is (10~\textmu m)$^2$, i.e. a $2500^2$ grid, with periodic boundary conditions in the lateral ($y$) direction and Neumann (no-flux) in the growth ($x$) direction.
The total simulated times ($t_{\rm f}$), listed in Table~\ref{tab:param:meltpool}, were verified to be sufficient to yield a steady microstructure behavior.

Each PF simulation was performed on one high-performance computing node.
Each node is equipped with 4 Nvidia RTX\,3090 GPUs and an AMD (16-core EPYC 7282) CPU processor.
The completion of each simulation lasted between 100~minutes (for $t_{\rm f}=0.04$\,ms at $V=1.0$\,\textmu m/s) and 16.5~hours ($t_{\rm f}=0.4$\,ms at $V=0.1$\,\textmu m/s).

%%%%%%%%%%%%%%%%%%%%%
\begin{table}[h]
\caption{Conditions of PF simulations along the melt pool tail.}
\label{tab:param:meltpool}
\centering
\footnotesize
\begin{tabular}{c | c c c c c c c c c c c c | l}
\br
$V$ & 0.1 & 0.2 & 0.3 & 0.4 & 0.5 & 0.6 & 0.7 & 0.8 & 0.9 & 0.95 & 0.99 & 1.0 & m/s\\
$G$ & 10.3 & 10.2 & 9.9 & 9.6 & 9.2 & 8.6 & 7.9 & 7.0 & 5.9 & 5.1 & 4.4 & 4.2 & K/\textmu m\\
$t_{\rm f}$ & 0.4 & 0.2 & 0.15 & 0.1 & 0.08 & 0.07 & 0.06 & 0.05 & 0.045 & 0.04 & 0.04 & 0.04 & ms	\\
\br
\end{tabular}
\end{table}
\normalsize
%%%%%%%%%%%%%%%%%%%%%

%%%%%%%%%%%%%%%%%%%%%%%%%%%%%%%%%%%%%%%%%%
\section{Results}
\label{sec:results}

\begin{figure}[h]
\centering
\includegraphics[width=.88\textwidth]{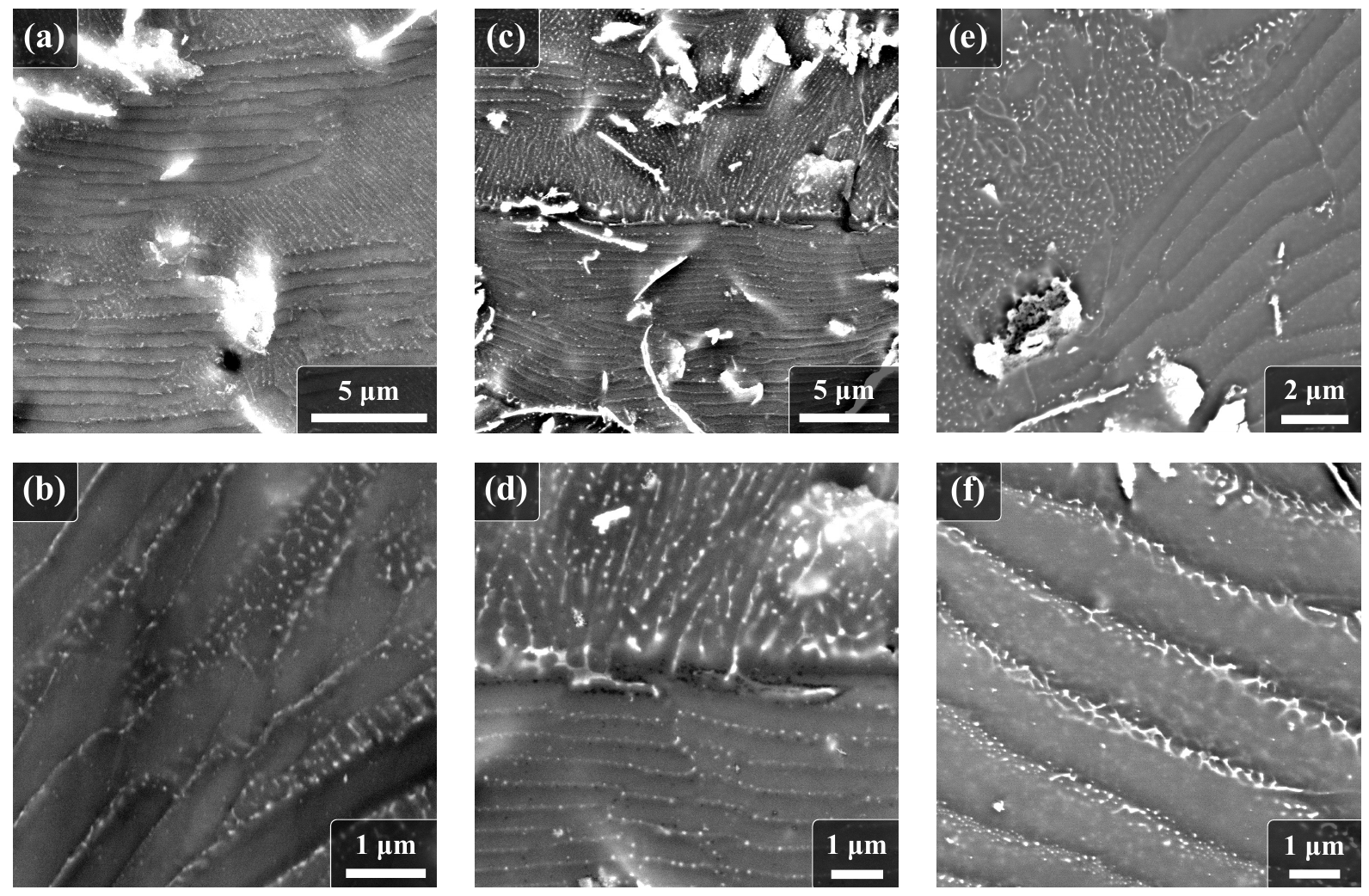}
\caption{Scanning electron microscope (SEM) images showing the microstructure of the WE43 Mg alloy processed by LPBF. 
The building direction is vertical for panels (a)-(d).
Panels (e)-(f) show surfaces normal to the build direction.
}
\label{Fig:SEM}
\end{figure}

\begin{figure}[h]
\centering
\includegraphics[width=.88\textwidth]{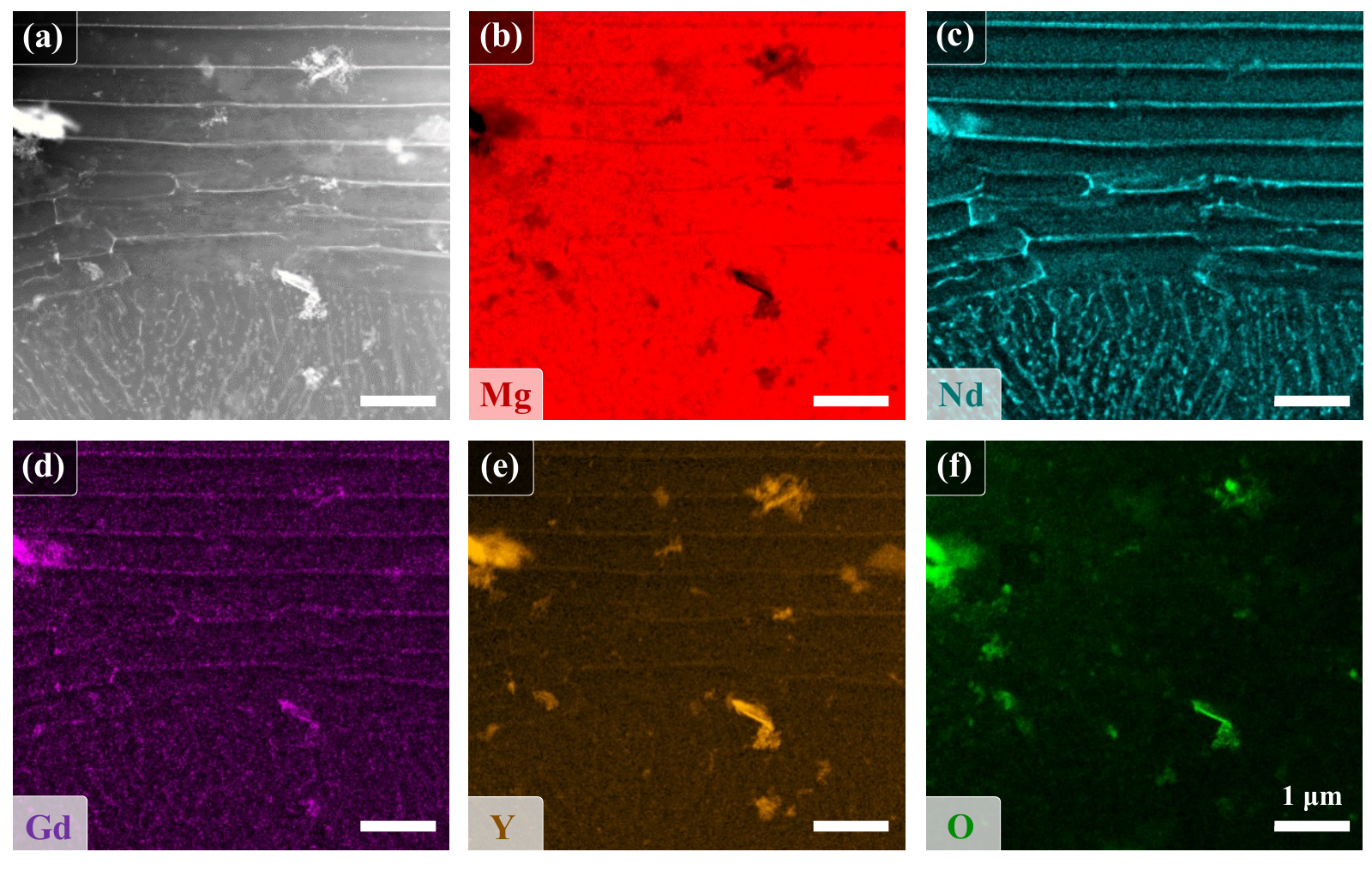}
\caption{
High-angle annular dark-field (HAADF) scanning transmission electron microscope (STEM) image of the microstructure in a transition region from dendrite/seaweed to banded microstructure (a) and (b-f) EDS mapping of chemical elements in region (a).
The building direction is vertical.
}
\label{Fig:EDX}
\end{figure}

Figure~\ref{Fig:SEM} shows SEM micrographs in various locations of the samples.
Figure~\ref{Fig:EDX} shows a TEM image, as well as corresponding compositional EDS maps for Mg, Nd, Gd, Y and O elements.

%%%%%%%%%%
%%%%%%%%%%

%%%%%%%%%%
%%%%%%%%%%

Bright irregular flakes measuring a few micrometers correspond to yttrium oxide particles, with small amount of Gd and Nd, but completely depleted of Mg (Figure~\ref{Fig:EDX}). These are inherited from the passivation layer of the feedstock powder, aimed to mitigate the high flammability of Mg. This high melting temperature oxide layer is fragmented during LPBF. Resulting flakes are transported by Mg fumes and end up mixed with the molten metal. Therefore, they do not directly participate in the solidification process during LPBF, besides acting as solid particles potentially interfering with the advance of the solidification front.

Within the regions solidified during LPBF (i.e. excluding the Y oxide flakes), micrographs show the co-existence of patterned (dendritic or seaweed) regions and banded regions, both of which are decorated with fine Mg-RE intermetallic precipitates (bright fine particles).
Images taken parallel to the build direction, upwards in figure~\ref{Fig:SEM}(a)-(d), mostly exhibit bands normal to the build direction, where one might expect a strong temperature gradient during AM.
Images showing a surface normal to the build direction, i.e. figure~\ref{Fig:SEM}(e)-(f), also exhibit both banded and non-banded regions.
Upon closer look into the banded structures (figure~\ref{Fig:SEM}(b) and (f)), the bright regions of the bands appear composed of fine patterns and intermetallic precipitates. 
The EDS compositional maps in Figure~\ref{Fig:EDX} confirm that bands are composed of chemically homogeneous dark regions alternating with bright regions particularly rich in Nd and also exhibiting significant segregation of Gd, and Y.

\begin{figure}[t]
\centering
\includegraphics[width=\textwidth]{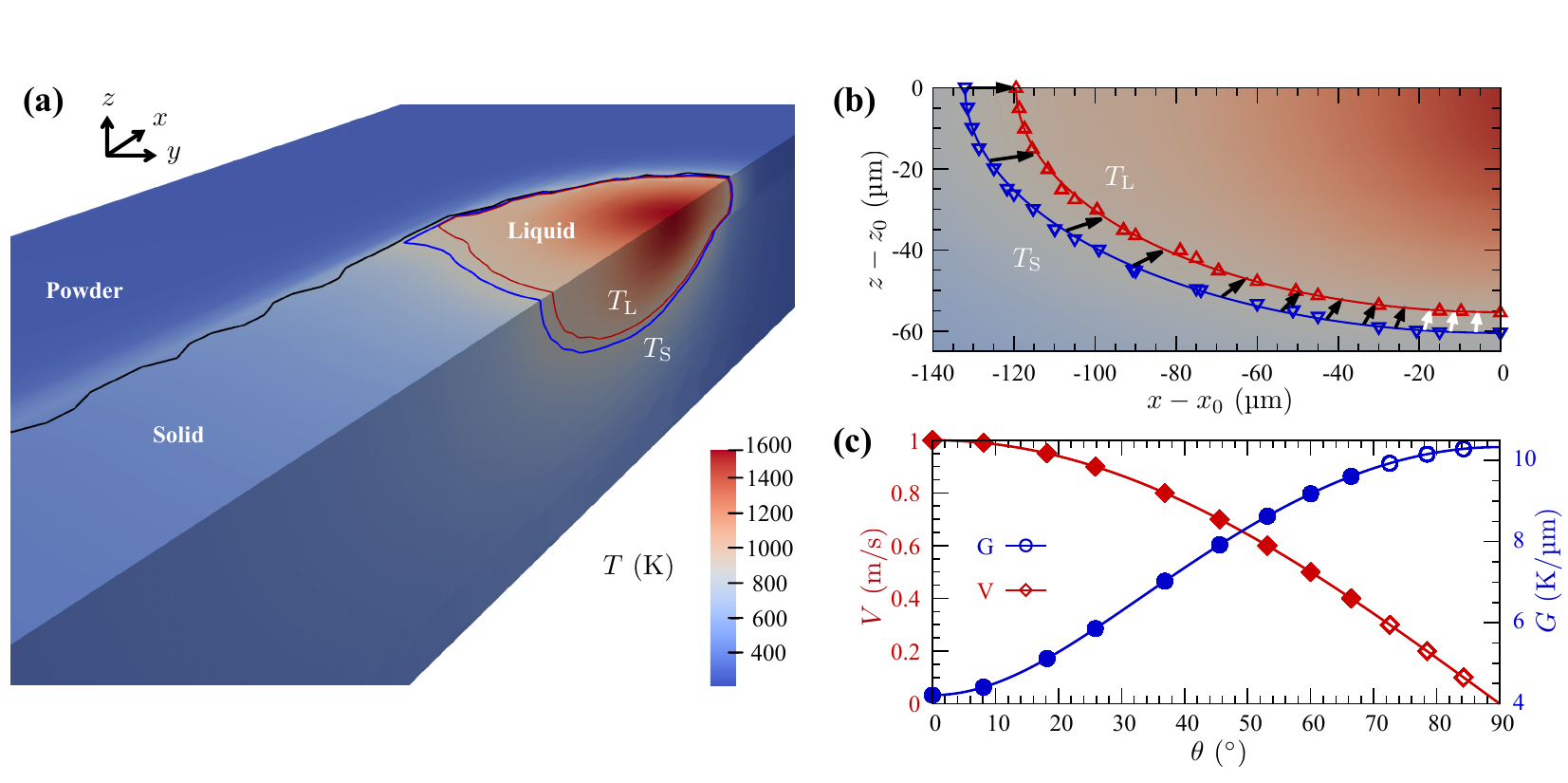}
\caption{
Results of the thermal FE simulation: (a) 3D temperature field (color map) in the upper $(y<y_{\rm L})$ region of the domain, (b) 2D temperature field (color map) in the tail region of the melt pool in the $(y=y_{\rm L})$ plane and location of the liquidus and solidus temperatures from the FE simulation (symbols) superimposed with their elliptical approximations \eref{eq:rl}-\eref{eq:rs} (lines), (c) solidification velocity \eref{eq:Vellip} and temperature gradient \eref{eq:Gellip} along the melt pool tail as a function of the polar angle $\theta$ \eref{eq:theta}.
Arrows in (b) and symbols in (c) illustrate $(G,V)$ conditions considered in the PF simulations (black arrows and filled symbols for cases resulting in an oscillatory banding instability, white arrows and open symbols otherwise -- see Figure~\ref{Fig:PFMeltPool}).
}
\label{Fig:FE}
\end{figure}

Figure~\ref{Fig:FE} illustrates the results of the FE simulation. 
Panel (b) shows superimposed location of the FE-predicted liquidus ($\vartriangle$ symbols) and solidus ($\triangledown$ symbols) isotherms and their elliptical approximations \eref{eq:rl}-\eref{eq:rs} (solid lines).
The melt pool dimensions are $d_{\rm L}=55.33$\,\textmu m, $d_{\rm S}=60.46$\,\textmu m, $l_{\rm L}=119.48$\,\textmu m, and $l_{\rm S}=132.06$\,\textmu m.
The resulting temperature gradient and solidification velocity along the melt pool, following equations \eref{eq:Gellip}-\eref{eq:Vellip} appear in figure~\ref{Fig:FE}(c).
The magnitude of the temperature gradient evolves from 10.3\,K/\textmu m at the bottom ($\theta=90^\circ$) to 4.2\,K/\textmu m at the tail ($\theta=0^\circ$) of the melt pool, whereas the solidification velocity jointly goes from 0 to $V_{\rm L}=1$\,m/s.
Arrows in figure~\ref{Fig:FE}(b) and symbols in figure~\ref{Fig:FE}(c) mark the location and $(G,V)$ conditions considered in the PF simulations.
Black arrows and filled symbols identify configurations leading to banded microstructures (otherwise, arrows are white and symbols are open), illustrating the predominance of the banding phenomenon in most of the melt pool.

%%%%%%%%%%
%%%%%%%%%%

\begin{figure}[t]
\centering
\includegraphics[width=\textwidth]{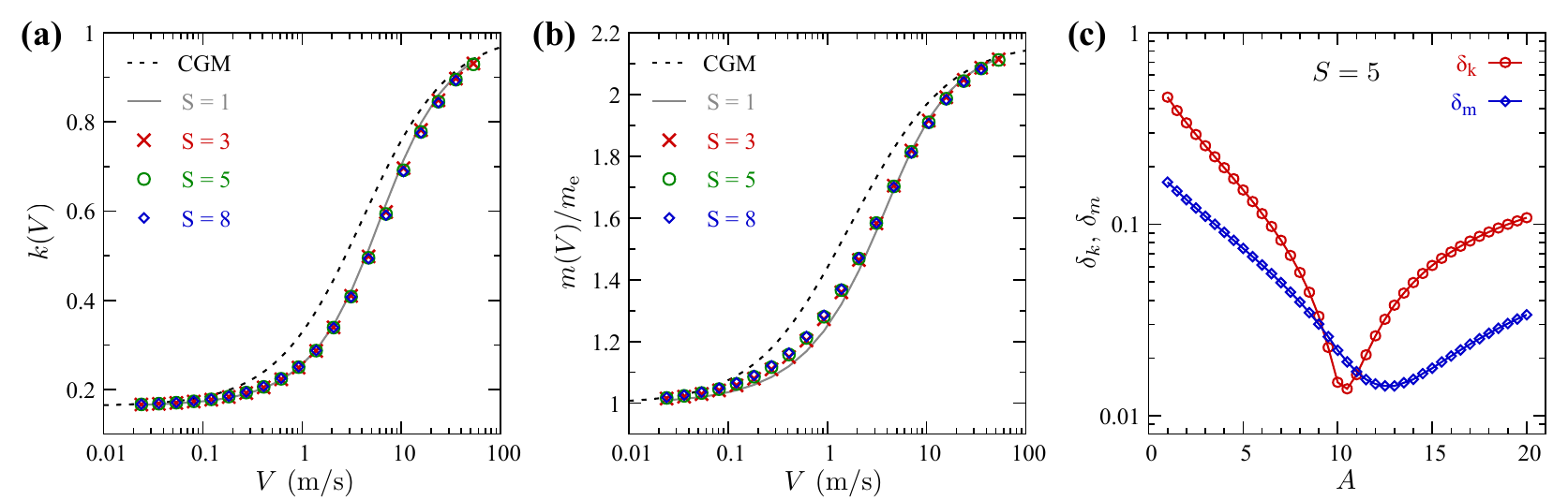}
\caption{
Results from preliminary PF calibration and verification steps, using a 1D formulation considering a steady-state profile $\phi_0(x)$ (see section~\ref{sec:meth:pf:simu}). 
Plots of (a) partition coefficient and (b) liquidus slope as a function of interface velocity for $(S,A)=(1,1)$ (solid gray line) and $(S,A)=(3,6)$, $(5,11)$, and $(8,18.5)$ (symbols) are compared to predictions of the CGM \eref{eq:cgm:kv}-\eref{eq:cgm:mv} (dashed black line).
(c) Calculated deviations $\delta_k$ and $\delta_m$ integrated over the entire $V$ range between PF-predicted $k(V)$ and $m(V)$ and the reference solution at $S=1$ and $A=1$, here illustrated for $S=5$.
}
\label{Fig:ConvergencePF}
\end{figure}

Figure~\ref{Fig:ConvergencePF} illustrates the results of the PF preliminary 1D convergence analysis. 
Panels (a) and (b) respectively show the velocity dependence of the interface solute partition coefficient, $k(V)$, and liquidus slope, $m(V)$, for different values of $(S,A)=(3,6)$, $(5,11)$, and $(8,18.5)$ (symbols), compared to the reference case with $(S,A)=(1,1)$ (solid gray line) and predictions of the CGM \eref{eq:cgm:kv}-\eref{eq:cgm:mv} (dashed black line). 
The latter was calculated considering a diffusion velocity through the interface $V_{\rm d} = 0.356\,V_{\rm d}^0\ln(1/k_{\rm e})/(1-k_{\rm e}) = 4.06$\,m/s with $V_{\rm d}^0 = D_{\rm L}/W_0 = 5.26$\,m/s, and a partial solute drag with $\alpha=0.645$ (see \cite{ji2023microstructural}).
The selected pairs of $(S,A)$ values were obtained by a minimization of $\delta_k(A)$ and $\delta_m(A)$ curves, illustrated for $S=5$ in figure~\ref{Fig:ConvergencePF}(c).

%%%%%%%%%%
%%%%%%%%%%

\begin{figure}[h]
\includegraphics[width=\textwidth]{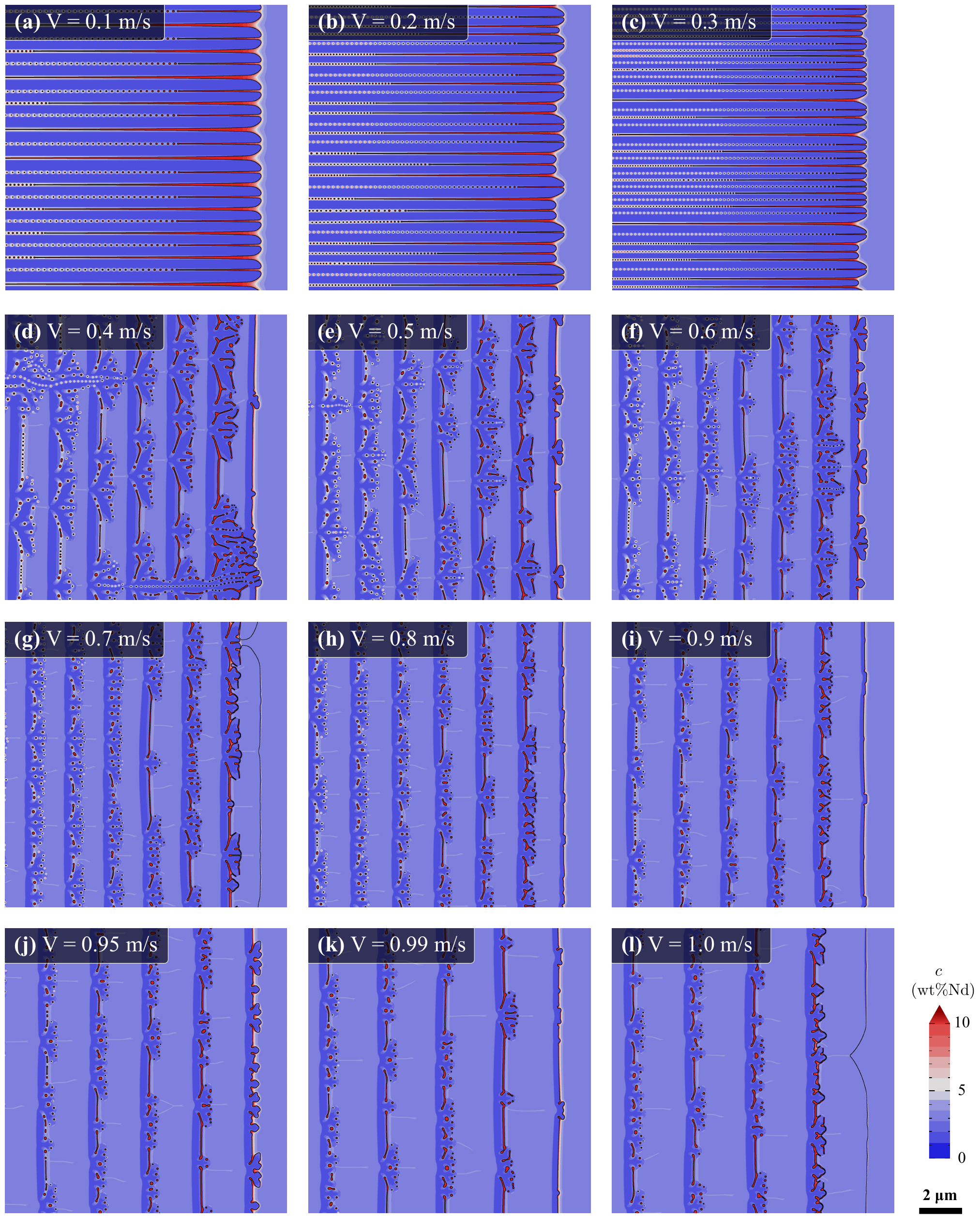}
\caption{
Final state, i.e. solid-liquid interface (black line) and solute concentration field (color map) at $t=t_{\rm f}$, in PF simulation for $(G,V)$ conditions of solidification from (a) the bottom to (l) the tail of the melt pool (see figure~\ref{Fig:PFMeltPool} and Table~\ref{tab:param:meltpool}).
The growth direction ($x+$) is represented from left to right.
(Video files for each panel are provided as supplementary material.)
}
\label{Fig:PFMeltPool}
\end{figure}

\begin{figure}[t]
\includegraphics[width=\textwidth]{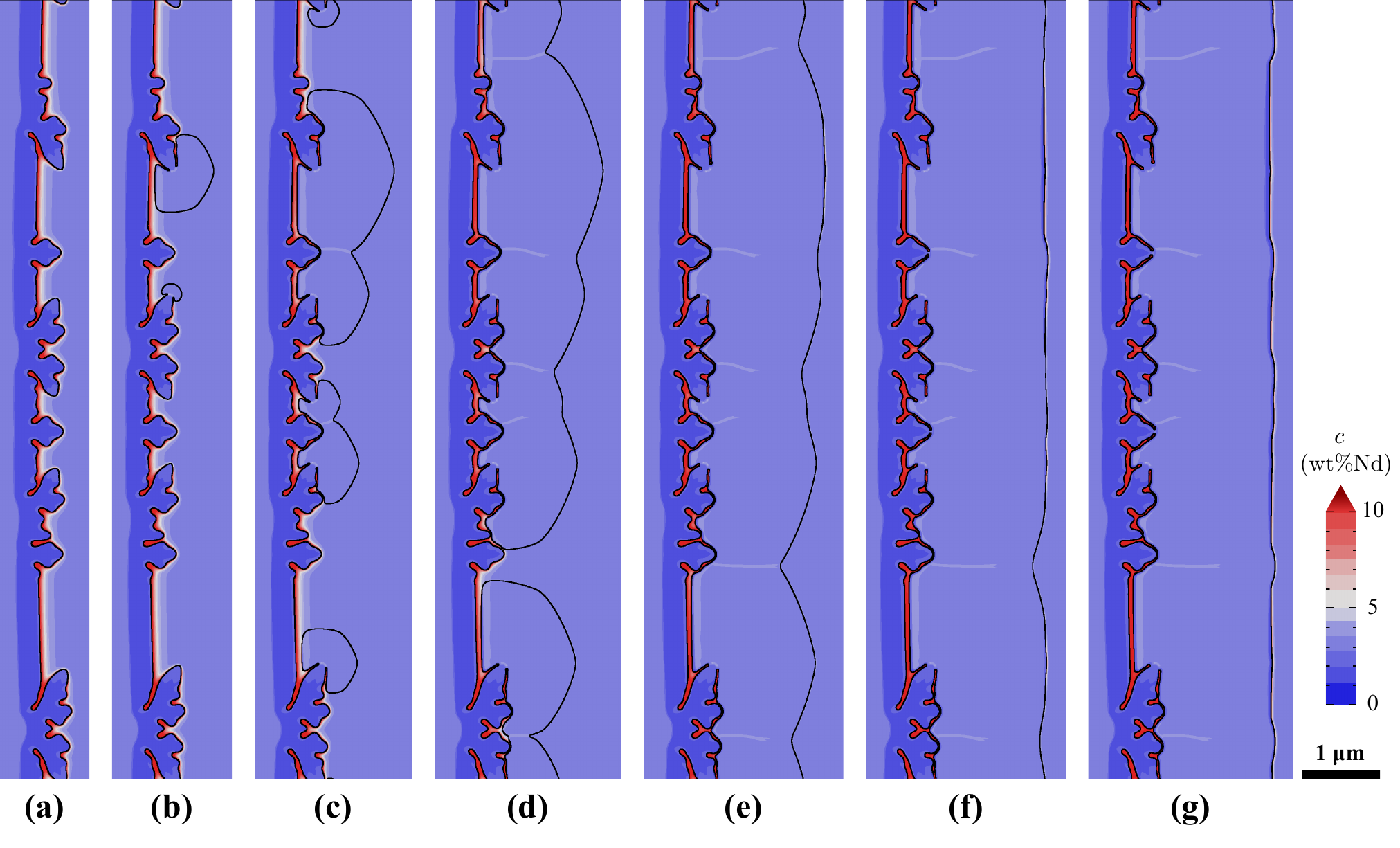}
\caption{
Evolution of the solid-liquid interface (black line) and solute concentration field (color map) during a banding cycle of the PF simulation at $V=1~$m/s and $G=4.2~$K/\textmu m (Figure~\ref{Fig:PFMeltPool}(l)).
The growth direction ($x+$) is represented from left to right.
Successive snapshots are 40\,nanoseconds apart.
}
\label{Fig:PFBurgeoning}
\end{figure}

Figures~\ref{Fig:PFMeltPool} and \ref{Fig:PFBurgeoning} show the results of the PF simulations, with a growth direction ($x+$) represented from left to right. 
Figure~\ref{Fig:PFMeltPool} illustrates the final microstructure obtained for $(G,V)$ conditions sampled along the melt pool tail (Table~\ref{tab:param:meltpool}).
Figure~\ref{Fig:PFBurgeoning} focuses on the interface region during one banding period for the case at $V=1~$m/s and $G=4.2~$K/\textmu m.
Both figures show the $(\phi=0)$ location of the interface (black line) superimposed on the solute concentration $c(x,y)$ color map.

%%%%%%%%%%
%%%%%%%%%%

%%%%%%%%%%
%%%%%%%%%%

%%%%%%%%%%%%%%%%%%%%%%%%%%%%%%%%%%%%%%%%%%
\section{Discussion}
\label{sec:discussion}

Microstructural characterization (figures~\ref{Fig:SEM} and \ref{Fig:EDX}) unambiguously identify the origin of banded regions as rapid solidification.
These microstructures are nearly identical to those reported from other rapid solidification processes \cite{kurz1996banded}, in particular in Al-Cu alloys \cite{carrard1992banded, zimmermann1991characterization, mckeown2014situ,mckeown2016time}.
They emerge from an alternation of microsegregation-free (partitionless) planar growth and dendritic (or seaweed) growth with solute partitioning.
Overall, banded microstructures are predominant and ubiquitous throughout our additively manufactured Mg alloy samples.
They do not appear to have been significantly affected by the thermal cycling due to the laser path after solidification. The as-printed microstructures (Figure~\ref{Fig:SEM}) retain features distinctive of rapid solidification banded microstructures, with solute segregation confined to the thin patterned regions  (Figure~\ref{Fig:EDX}). Our interpretation is that the thermal peaks induced by the laser are too fast to trigger any significant diffusion mechanism in the solidified microstructure. However, after a possible post-printing heat treatment, we expect that the fine precipitate, present in both dendritic and banded grain could be fully dissolved into the matrix (but not the Y oxides) \cite{li2021microstructure}.
The predominance of banded microstructures is consistent with the results of thermal and microstructure selection simulations, indicating that the majority of the melt pool (black arrows in figure~\ref{Fig:FE}(b)) solidifies under conditions promoting a banding instability, here found for $V\geq0.4$\,m/s (figure~\ref{Fig:PFMeltPool}).

The PF convergence study confirms that, for this alloy, PF simulations retain a good accuracy on both $k(V)$ (figure~\ref{Fig:ConvergencePF}(a)) and $m(V)$ (figure~\ref{Fig:ConvergencePF}(b)) over the entire range of considered interface velocity.
As mentioned in \cite{ji2023microstructural}, while the asymptotic limits at $V\ll V_{\rm d}$ and $V\gg V_{\rm d}$ match the CGM predictions, a small deviation appears at intermediate velocities, which is expected when comparing a diffuse interface approach (PF) with a sharp-interface theory (CGM).
Still, the close matching of both $k(V)$ and $m(V)$ curves for $S>1$ with the reference case at $S=1$ shows that the interface may be upscaled by up to a factor 8 with negligible deviation, hence ensuring that the calculations performed here with $S=5$ and $A=11$ remain very close to those at $S=1$. 
In 2D with an explicit time stepping scheme, the grid coarsening by a factor 5 leads to a reduction of computational cost by $\approx5^4=625$.
Results for $S=8$ even suggest that further upscaling by $8^4=4096$ is reasonable, which could possibly allow quantitative simulations at larger scale, e.g. possibly up to the full melt pool scale \cite{elahi2022multiscale}.

The mechanism leading to banded microstructures, illustrated in figure~\ref{Fig:PFBurgeoning}, is the burgeoning instability described in \cite{ji2023microstructural}, whereby dendritic tips accelerate, morphologically destabilize, and grow almost (yet not completely) isotropically, until they form a new planar front while decelerating. 
The sudden tip acceleration and morphological destabilization coincides with nearly complete solute trapping, as the solid phase appears to grow at the same solute concentration as the surrounding liquid.
Within its basal plane, the hcp Mg crystal has six $\langle11\bar20\rangle$ preferred growth directions, one of which is aligned with the temperature gradient direction, resulting in secondary sidebranches growing at a $\pm60^\circ$ angle from the main ($x+$) growth direction.
Branches growing at a $60^\circ$ angle with respect to the $x+$ direction are almost systematically at the origin of the burgeoning instability. 
This may be attributed to their relatively higher growth velocity in their specific growth direction, effectively faster due to the $60^\circ$ angle, and hence triggering the instability earlier than dendritic tips oriented along the temperature gradient direction.
PF results show that the dendrite acceleration first leads to significant solute trapping while still dendritic (indicated by the lower shade of blue of dendrite tips in figure~\ref{Fig:PFBurgeoning}(a)-(b)) before the onset of the burgeoning interface instability.
The instability arises almost simultaneously from several dendritic tips (figure~\ref{Fig:PFBurgeoning}(b)-(c)).
The locations where different partitionless regions impinge upon one another form slightly segregated channels, and a small deviation from a planar front (figure~\ref{Fig:PFBurgeoning}(e)-(g)), which often corresponds to the location of the emerging dendrites in the next banding cycle (see attached videos).

The effect of the growth velocity $V$ is apparent in both dendritic and banding regimes (figure~\ref{Fig:PFMeltPool}).
In the dendritic regime ($V\leq0.3$\,m/s), the interface is mostly composed of dendritic doublon structures, with a notable decrease of primary spacing as $V$ increases.
Doublons consist of two asymmetric fingers, with a mirror symmetry between them, growing cooperatively with a narrow liquid channel running between the two fingers along the symmetry axis.
They constitute a known steady-state growth solution promoted by low anisotropy and high undercooling \cite{brener1992kinetic, ihle1994fractal, amar1995parity, akamatsu1995symmetry, kupferman1995coexistence}.
Theoretical and computational studies \cite{brener1992kinetic, ihle1994fractal, amar1995parity, brener1996cellular} of the resulting sharp-interface problem have shown that, when both doublon and dendrite steady-state solutions exists, doublons grow faster than regular symmetric dendrites, thus giving the doublon an advantage in terms of microstructure selection.
For low interface anisotropy, doublons have been described as the “building block” of seaweed structures \cite{brener1996structure}. However, the transition from dendrite to seaweed was shown to occur without the formation of doublon or multiplet when the interface dynamics is strongly dominated by kinetics \cite{bragard2002linking}.
Here, the propensity to form doublons may be attributed to a combination of (i) the low interface energy anisotropy, (ii) the 2D simulations (doublons were mostly observed in 2D thin sample experiments \cite{akamatsu1995symmetry}, while 3D calculations \cite{abel1997three,debierre2016phase} and experiments \cite{singer2009experimental} demonstrated the possibility of more complex multiplet structures), and (iii) the high temperature gradient $G$. 
Further PF calculations on a broader range of $G$ (to be reported elsewhere) suggest that a transition toward a seaweed pattern occurs between 10 and 50\,K/\textmu m, which seems consistent with the substitution of symmetric dendrites by doublons at $G\approx10$\,K/\textmu m.
The transition between dendritic and seaweed growth modes was shown to be discontinuous (i.e. first order), with a range of parameters where both branches of solutions may be (meta)stable \cite{ihle1994fractal}. It was also shown that seaweed patterns are promoted by a high temperature gradient \cite{tourret2015growth, utter2001alternating, provatas2003seaweed}. 
The transition to the seaweed regime and the presence of doublons might also be promoted by the sixfold anisotropy, although this hypothesis remains to be studied in further depth.

Our results suggest a transition velocity between dendrites/doublons and banded patterns between 0.3 and 0.4\,m/s.
This is lower than the absolute stability predicted by linear stability analysis \eref{eq:absstab}, $V_{\rm a}\approx 2.41$\,m/s using $k(V)$ and $m(V)$ from figure~\ref{Fig:ConvergencePF} (i.e., for simplicity, using the 1D approximation $\phi(x)\approx\phi_0(x)$ and directly integrating \eref{eq:dcdt_1D_steady}-\eref{eq:mV_1D_steady}) or $V_{\rm a}\approx 1.85$\,m/s using the CGM \eref{eq:cgm:kv}-\eref{eq:cgm:mv}.
However, it is worth noting that the precise maximum stable velocity for a steady dendrite (or doublon) can only be probed by first stabilizing a dendritic pattern, before slowly increasing the $V$ until its morphological destabilization (see \cite{ji2023microstructural}), and that the resulting threshold $V$ is expected to be highly sensitive to fluctuations/noise. 
Yet, here we start with a noisy quasi-planar interface, which appears more likely to yield an oscillatory instability rather than dendrites or doublons across a velocity range where both steady-state solutions may effectively exist.
As such, the value identified here ($0.3\sim0.4$\,m/s) can only be viewed as a lower bound for the maximum dendrite/doublon steady velocity, which may still reasonably approximate non-ideal (noisy) conditions in experiments.
Still, while the order of magnitude is realistic, a deeper study on the effect of noise (and a possible $V$-ramp approach as in \cite{ji2023microstructural}) would be required to definitely conclude on the maximum steady dendrite/doublon velocity for this alloy.

At $V=0.4$\,m/s, while most of the domain undergoes the cyclic banding instability, some dendritic/doublon patterns may coexist and grow jointly with the bands during a few (here, 2 or 3) successive banding cycles (see top left and bottom right of figure~\ref{Fig:PFMeltPool}(d) and corresponding attached video).
In the banding regime ($V>0.4$\,m/s), an increase of growth velocity leads to a reduction of the width of the patterned/segregated region, consistent with experimental observations in rapidly solidified Al alloys \cite{kurz1996banded,gremaud1991banding}.

PF simulations (figure~\ref{Fig:PFMeltPool}) predict the emergence of microstructures qualitatively similar to those observed in experimental micrographs (figure~\ref{Fig:SEM}).
Since the selected PF model considers only the formation (solidification) of a primary phase, simulation results show the expected solid-liquid interface pattern, discarding the formation of secondary solid phases that would nucleate in high-Nd segregated liquid regions.
The quantitative comparison between experiments and simulated microstructures is reasonable, with a band spacing of about one to two micrometers.
However, some experimental band spacings (e.g. in figure~\ref{Fig:SEM}(a),(c)) appear lower than predicted by PF simulations.
This discrepancy may be attributed to two main factors. 
First, the fact that simulations are two-dimensional, while the experiments are inherently 3D, with a much more complex temperature history than what we considered here, and the fact that micrographs are only representing a 2D slice of the resulting three-dimensional banded microstructures.
Second, and importantly, the fact current PF simulations neglect the effect of the latent heat release, which was shown to have a central role in the selection of the band spacing \cite{karma1992dynamics,karma1993interface,conti1998heat,ji2023microstructural}.
The latent heat was also shown to potentially induce a lateral spreading of the interface instability, resulting in microsegregation-free bands slightly tilted with respect to the temperature gradient \cite{ji2023microstructural}, which might also affect the near simultaneous burgeoning instability observed throughout the sample in the present simulations.
A deeper investigation on the effect of the latent heat is currently underway and will be reported elsewhere.
Other possible sources of discrepancies could stem from high uncertainties on material parameters, such as the liquid diffusion parameter, here roughly estimated as $5.26\times10^{-9}~$m$^2$/s, which seems high but of the typical order of magnitude for liquid metallic alloys ($\mathcal{O}(10^{-9})$\,m$^2$/s).
The high flow velocities expected in the melt pool may also play a role, in particular resulting in distinct local solidification conditions (e.g. solute concentration, temperature gradient). 
However, given the small volume of the melt pool and the typically low diffusion length for such high solidification rates, we expect that, at the scale of our simulations, the assumption of nearly flat thermal field and a homogeneous (well stirred) liquid remain reasonable.

%%%%%%%%%%%%%%%%%%%%%%%%%%%%%%%%%%%%%%%%%%
\section{Summary \& Perspectives}
\label{sec:conclu}

We studied the formation of rapid solidification microstructures in 3D-printed (LPBF) samples of biomedical-grade Mg alloy (WE43), combining advanced characterization, finite element thermal simulation at the laser track scale, and phase-field simulations of rapid solidification at the scale of the solid-liquid interface pattern.

Our experiments confirm the prevalence of banded structures composed of pattern-free microsegregation-free regions alternating with patterned (dendrite or seaweed) regions with solute segregation and intermetallic precipitates. 
These microstructures are clearly indicative of a rapid solidification regime at an intermediate velocity $V$ between dendritic and planar growth, previously reported in many rapidly solidified alloys \cite{kurz1996banded} --- but, to the best of our knowledge, never discussed so far in additively manufactured samples.
Our simulations confirm that the LPBF range of solidification conditions, namely temperature gradient $G$ and growth velocity $V$, promote the occurrence of banded microstructures throughout the majority of the melt pool (except at its very bottom where the local $V\leq 0.3$\,m/s).

The underlying mechanism of the banding phenomenon was confirmed to be the burgeoning instability of primary (or secondary) dendrite tips recently revealed via PF simulations of rapid solidification in thin Al-Cu TEM samples \cite{ji2023microstructural}.
For a Mg crystal with sixfold anisotropy in the basal plane, this instability originates principally from sidebranches, growing at a $60^\circ$ angle from the primary ($x+$) growth direction, due to their relatively higher velocity respective to their growth direction.
The interface pattern at a $V$ just below the onset of the banding instability exhibit a high density of doublons, attributed to the simulation being two-dimensional, the low capillary anisotropy, and the high temperature gradient $G$.
As expected, an increase of $V$ leads to a reduction of primary spacing in the dendritic/doublons regime, and to a reduction of the width of the regions exhibiting segregation and microstructural features in the banded growth regime. 
The effect of $G$ was not analyzed in further details, since our PF simulations do not explicitly include the effect of latent heat release and diffusion during solidification, which is known to have a dominant effect over that of the higher-scale temperature gradient \cite{karma1993interface,ji2023microstructural}. 
In spite of this limitation, the present study clearly demonstrates the potential of physics-based simulations in predicting the emergence of rapid solidification microstructures, particularly in the context of metal AM.

Perspectives from this work are multiple. 
The most direct follow-up (and focus of ongoing work), is the incorporation of the latent heat  in the PF model, as well as a deeper study into its effect on the selection of microstructure.
The main associated challenge comes from the different time scales of heat and solute transport, which needs to be addressed computationally, for instance using multiple spatial grids \cite{ji2023microstructural,song2018thermal}.
Another important contribution would be the study of the burgeoning instability and banding mechanism in three dimensions. 
Our results suggest that, with the present PF model, the diffuse interface width $W$ might possibly be coarsened even further than the value $S=5$ used here and in previous works \cite{ji2023microstructural} while retaining a reasonable quantitative accuracy on $k(V)$ and $m(V)$.
Therefore, although it would certainly represent a substantial computational effort, three-dimensional simulations might be within reach, considering the ever-increasing size and speed of high-performance computing hardware (in particular GPUs).

Finally, from a longer-term applied viewpoint, as rapidly solidified Mg alloy were reported to exhibit significantly different mechanical \cite{gjestland1991stress, srivatsan1995tensile, zhang2013improving} and corrosion \cite{liao2012improved,zhang2013improving, willbold2013biocompatibility, shuai2017laser} behavior than regular casting products, one may expect that locally changing the solidification conditions could enable a local modification of microstructures and properties, and ultimately performance of additively manufactured Mg parts.
Combining microstructural models with track-scale or macroscopic thermal models of additive manufacturing could provide a greater understanding of the effect of various parameters (e.g. laser power/speed/path, part geometry, alloy chemistry) on the local distribution of microstructures within a part, e.g. an implant.
Yet, while rapidly solidified Mg alloys may exhibit enhanced mechanical properties and corrosion resistance, a deeper fundamental understanding at microstructural level remain lacking (e.g., to the best of our knowledge, the effect of banded microstructures on corrosion properties is not clearly established).
Once both process-microstructure and microstructure-properties links are clearly established, one may envision computational strategies based on optimization algorithms, not only for topology, but also for alloy chemistry and local processing parameters in order to achieve localized optimal properties.
In this context, we trust that the quantitative prediction of microstructures and microstructural length scales, and their correlation with local solidification conditions, may play a key role in guiding the design of novel 3D-printed biomedical implants with locally tuned mechanical and corrosion properties.

%%%%%%%%%%%%%%%%%%%%%%%%%%%%%%%%%%%%%%%%%%
\ack

The authors wish to thank Alain Karma and Kaihua Ji for enlightening discussions and clarifications regarding the phase-field model. 
DT, RT, and AKB gratefully acknowledge support from the Spanish Ministry of Science and the European Union NextGenerationEU, through a Ram\'on y Cajal Fellowship [RYC2019-028233-I] (DT), the MiMMoSA project [PCI2021-122023-2B] (RT), and the HexaGB project [RTI2018-098245] (AKB).
ADB acknowledges financial support from the European Commission, Horizon Europe, Marie Skłodowska-Curie Actions through the M3TiAM project [Grant agreement 101063099].
ML and JMA acknowledge support by the European Union’s Horizon Europe research and innovation programme [Grant agreement 101047008], the Spanish Government [PID2019-109962RB-100 and PID2022-138417OB-C21], and Madrid Regional Government [Y2020/BIO-6756: i-MPLANTS-CM].

%%%%%%%%%%%%%%%%%%%%%%%%%%%%%%%%%%%%%%%%%%
\appendix
\section*{Supplementary Material}

The joint multimedia files contains videos corresponding to each panel of figure~\ref{Fig:PFMeltPool}.
For cases leading to banding instability (i.e. for $V\geq0.4$\,m/s) the last 10\% of the simulation time is displayed with a lower time step between frames (1/1000 of the simulated duration, instead of 1/100 for the first 90\%) in order to reveal temporal details of the fast growth dynamics during banding.

\section*{Appendix: Developed expressions of anisotropic terms}
\label{sec:appdx}
\setcounter{section}{1}
The anisotropic excess free energy of the solid-liquid interface is expressed as
\begin{equation}
\gamma(\theta) = \gamma_0 a_{\rm s}(\theta) 
\quad.
\end{equation}
We simplify its notations and that of its derivatives with respect to $\theta$, simply writing $a_{\rm s}$ instead of $a_{\rm s}(\theta)$. 
Hence, from \eref{eq:aniso_g}, we have
\begin{eqnarray}
a_{\rm s} &= 1+ a_{20} + a_{66}\cos(6\theta) \\
a_{\rm s}' &= \partial_\theta a_{\rm s} = -6 a_{66}\sin(6\theta) \\
a_{\rm s}'' &= \partial_{\theta\theta} a_{\rm s} =-36 a_{66}\cos(6\theta)
\end{eqnarray}
with
\begin{eqnarray}
a_{20} &= -\frac{\varepsilon^\gamma_{20}}{4}\sqrt{\frac{5}{\pi}} \\
a_{66} &= \frac{\varepsilon^\gamma_{66}}{64}\sqrt{\frac{6006}{\pi}}
\quad.
\end{eqnarray}
The function $a_{\rm s}$ and its derivatives can be expressed directly as a function first order spatial derivatives of the phase field $\phi$ using trigonometric identities
\begin{eqnarray}
\cos(6\theta) &= (2\cos^2\theta -1)\left[ 4(2\cos^2\theta -1)^2-3 \right] \\
\sin(6\theta) &= 2\sin\theta\cos\theta \, (3-16\sin^2\theta\cos^2\theta)
\end{eqnarray}
with 
\begin{eqnarray}
\cos\theta &= \frac{\partial_x\phi}{|\vec\nabla\phi|} \\
\sin\theta &= \frac{\partial_y\phi}{|\vec\nabla\phi|}
\quad.
\end{eqnarray}
Then, following the same method as in \cite{tourret2015growth} (Appendices therein), the first term on the right-hand side of \eref{eq:dpdt} may be expanded as 
\begin{equation}
\vec\nabla \cdot\left[ W(\mathbf{n})^2\vec\nabla \phi\right] = 
S^2W_0^2 \left[ \vec\nabla(a_{\rm s}^2) \cdot\vec\nabla\phi + a_{\rm s}^2\nabla^2\phi \right]
\end{equation}
with the first term in brackets expressed as
\begin{equation}
\label{eq:a12}
\vec\nabla(a_{\rm s}^2) \cdot\vec\nabla\phi 
= (\partial_x\phi\partial_x\theta + \partial_y\phi\partial_y\theta )2 a_{\rm s}'a_{\rm s}
\quad.
\end{equation}
Moreover, the penultimate term on the right-hand side of \eref{eq:dpdt} may be expressed as 
\begin{equation}
\label{eq:a13}
\sum_{\eta=x,y}\left[ \partial_\eta \left( |\vec\nabla\phi|^2 W(\mathbf{n}) \frac{\partial W(\mathbf{n})}{\partial (\partial_\eta\phi)}\right) \right] 
= (\partial_x\phi\partial_y\theta - \partial_y\phi\partial_x\theta )(a_{\rm s}''a_{\rm s}+a_{\rm s}'^2)
~~,
\end{equation}
where, since $\theta=\tan^{-1}\left(\partial_y\phi/\partial_x\phi \right)$, we have \cite{tourret2015growth} 
\begin{eqnarray}
\partial_x\theta &= \frac{\partial_{xy}\phi\partial_{x}\phi-\partial_{xx}\phi\partial_{y}\phi}{|\vec\nabla\phi|^2} \\
\partial_y\theta &= \frac{\partial_{yy}\phi\partial_{x}\phi-\partial_{xy}\phi\partial_{y}\phi}{|\vec\nabla\phi|^2}
\quad.
\end{eqnarray}
Thus, all terms in \eref{eq:a12} and \eref{eq:a13}, and hence all terms in \eref{eq:dpdt} may be expressed as local first and second order spatial derivatives of $\phi$.

%%%%%%%%%%%%%%%%%%%%%%%%%%%%%%%%%%%%%%%%%%
\section*{References} 
\bibliographystyle{iopart-num}
\bibliography{References}

\end{document}